\documentclass[12pt,letterpaper]{article}
\usepackage{amsmath,amssymb,array,calc,rotating,epsfig,psfrag,amscd}

\usepackage{bm,amsfonts,array,calc,rotating}
\usepackage{epsfig}
\usepackage{graphicx}
\usepackage{color}
\usepackage{slashed}
\usepackage[colorlinks=false]{hyperref}

\global\def\draftcontrol{0}

   \def\versionno{Higher Derivative Brane Couplings from String Amplitudes}

\catcode`\@=11

\expandafter\ifx\csname
draftcontrol\endcsname\relax\global\def\draftcontrol{0} \fi

{\count255=\time\divide\count255 by 60
\xdef\hourmin{\number\count255} \multiply\count255
by-60\advance\count255 by\time
\xdef\hourmin{\hourmin:\ifnum\count255<10 0\fi\the\count255}}
\def\draftdate{\number\month/\number\day/\number\year\ \ \ \hourmin }

\newcommand\makepapertitle{\par

  \begingroup
    \renewcommand\thefootnote{\@fnsymbol\c@footnote}%
    \def\@makefnmark{\rlap{\@textsuperscript{\normalfont\@thefnmark}}}%
    \long\def\@makefntext##1{\parindent 1em\noindent
            \hb@xt@1.8em{%
                \hss\@textsuperscript{\normalfont\@thefnmark}}##1}%
     \newpage
     \global\@topnum\z@   
     \@makepapertitle
     \thispagestyle{empty}\@thanks
  \endgroup
  \setcounter{footnote}{0}%
  \global\let\thanks\relax
  \global\let\makepapertitle\relax
  \global\let\@makepapertitle\relax
  \global\let\@thanks\@empty
  \global\let\@author\@empty
  \global\let\@date\@empty
  \global\let\@title\@empty
  \global\let\title\relax
  \global\let\author\relax
  \global\let\date\relax
  \global\let\and\relax
  \def\version{\let\version\@version\@gobble}
}
\def\@makepapertitle{%
  \newpage
   \ifnum\draftcontrol=1{}
   \version\versionno
   \vskip 5em%
   \else
   \hfill\hbox to 3cm {\parbox{4cm}{\@pubnum}\hss}%
   \vskip 5em%
   \fi
   \begin{center}%
   \let \footnote \thanks
      {\hskip -0\textwidth \hbox to 1\textwidth%
        {\centerline{\Large\bf{\noindent\@title}}}}%
     \vskip 2em%
     {\normalsize
       \lineskip .5em%
       \begin{tabular}[t]{c}%
         \@author
       \end{tabular}\par}%
     \vskip 1em%
     {\@bstract}%
     \end{center}%
     \vfill
     \@date%
     \vskip 1.5em%
     \noindent
     \rule{12em}{.02em}\par\noindent
     \@email%
   \par
}

\gdef\@pubnum{}
\def\pubnum#1{%
  \gdef\@pubnum{#1}}

\gdef\@bstract{}
\def\Abstract#1{%
  \gdef\@bstract{%
   \parbox{\textwidth-0pc}{%
   \centerline{\bf Abstract}\penalty1000
   \noindent
   \renewcommand\baselinestretch{1.0}
   {#1}}}
}

\gdef\@email{}
\def\email#1{%
   \gdef\@email{%
  {\small Email: {\tt #1}}}
}

\def\ps@paper{\let\@mkboth\@gobbletwo%
     \ifnum\draftcontrol=1
        \def\@oddfoot{\hbox to \textwidth{\tiny \versionno \hfil\tiny\draftdate}%
        \hskip -\textwidth \hbox to \textwidth{\hfil\rm\thepage\hfil}}%
     \else\def\@oddfoot{\hbox to \textwidth{\hfil\rm\thepage\hfil}}
     \fi
     \let\@evenfoot\@oddfoot
}

\def\body{\clearpage
          \pagestyle{paper}
        }
\newenvironment{acknowledgments}{%
\vskip 3.25ex
\noindent {\bf Acknowledgments}
}

\def\@version#1{\ifnum\draftcontrol=1
\typeout{}\typeout{#1}\typeout{}
\vskip3mm\centerline{\hbox{\fbox{\normalsize{\tt DRAFT -- #1 -- }
                   {\draftdate}}}}\vskip3mm
\fi}
\let\version\@version
\long\def\eqlabel#1{\ifnum\draftcontrol=1
                    \tag@false  
                    \tag*{(\theequation) \hbox to -0.2cm{\hspace{0cm}\small{#1}\hss}}
                    \refstepcounter{equation}
                    \edef\@currentlabel{\theequation}
                    \ltx@label{#1}          
                    \else
                    \label{#1}
                    \fi
                    }
\let\st@bibitem\@bibitem
\let\st@lbibitem\@lbibitem
\ifnum\draftcontrol=1
  \def\@bibitem#1{%
    \st@bibitem{#1}\a@@label{#1}\ignorespaces}
  \def\@lbibitem[#1]#2{%
    \st@lbibitem[#1]{#2}\a@@label{#2}\ignorespaces}
  \def\a@@label#1{%
    \gdef\a@lab{\smash{\normalfont\small#1}}
    \ifvmode
      \if@inlabel
        \global\setbox\@labels\hbox{%
          \llap{\a@lab\let\a@lab\relax
                \kern\@totalleftmargin\kern\marginparsep}%
          \box\@labels}%
      \fi
    \fi}
\fi

\renewcommand\baselinestretch{1.25}
\numberwithin{equation}{section}

\long\def\@makecaption#1#2{%
  \vskip\abovecaptionskip
  \sbox\@tempboxa{{\bf #1:} #2}%
  \ifdim \wd\@tempboxa >\hsize
    {\small\bf #1:} {\small #2}\par
  \else
    \global \@minipagefalse
    \hb@xt@\hsize{\hfil\box\@tempboxa\hfil}%
  \fi
  \vskip\belowcaptionskip}

\definecolor{refcol}{rgb}{0.2,0.2,0.8}
\definecolor{eqcol}{rgb}{.6,0,0}
\definecolor{purple}{cmyk}{0,1,0,0}

\gdef\@citecolor{refcol} \gdef\@linkcolor{eqcol}
\def\colorlinkspurple{\gdef\@urlcolor{purple}}
\def\colorlinksblue{\gdef\@urlcolor{blue}}
\def\colorlinksred{\gdef\@urlcolor{red}}

\def\revise#1       {\raisebox{-0em}{\rule{3pt}{1em}}%
                     \marginpar{\raisebox{.5em}{\vrule width3pt\
                     \vrule width0pt height 0pt depth0.5em
                     \hbox to 0cm{\hspace{0cm}{%
                     \parbox[t]{4em}{\raggedright\footnotesize{#1}}}\hss}}}}

\newcommand{\nc}{\newcommand}

\def\a{\alpha}

\def\d{\delta}
\def\e{\epsilon}

\def\th{\theta}

\def\m{\mu}
\def\n{\nu}
\def\x{\xi}

\def\G{\Gamma}
\def\D{\Delta}

\def\L{\Lambda}

\def\o{\over}
\def\pa{\partial}

\nc{\BB}{{\mathbb B}}
\nc{\CC}{{\mathbb C}}
\nc{\DD}{{\mathbb D}}
\nc{\EE}{{\mathbb E}}
\nc{\FF}{{\mathbb F}}
\nc{\GG}{{\mathbb G}}
\nc{\HH}{{\mathbb H}}
\nc{\JJ}{{\mathbb J}}
\nc{\RR}{{\mathbb R}}
\nc{\PP}{{\mathbb P}}
\nc{\QQ}{{\mathbb Q}}
\nc{\ZZ}{{\mathbb Z}}
\nc{\CP}{{\CC\PP}}
\nc{\calone}{{\mathbb 1}}

\newcommand{\C}[1]{$(\ref{#1})$}

\def\tr{{\rm tr}}

\newcommand{\lp}{\left(}
\newcommand{\rp}{\right)}
\newcommand{\ls}{\left[}
\newcommand{\rs}{\right]}

\newcommand{\hlf}{\frac{1}{2}}
\newcommand{\be}{\begin{equation}}
\newcommand{\ee}{\end{equation}}

\newcommand{\bea}{\begin{eqnarray}}
\newcommand{\eea}{\end{eqnarray}}

\begin{document}



\title{Four-Derivative Brane Couplings from String Amplitudes}
\date{October 17, 2011}

\author{\\[.1cm]Katrin Becker$^{1,2}$, Guangyu Guo$^{1,3}$ and Daniel Robbins$^{1,2}$\\
\\[.4cm]$^1$ \it Department of Physics, Texas A\&M University, \\ \it College Station, TX 77843, USA
\\ [.2cm]  $^2$ \it School of Natural Science, Institute for Advanced Study,\\ \it Einstein Drive, Princeton, NJ 08540, USA
\\[.2cm] $^3$ \it Mathematical Sciences Certer, \\ \it Tsinghua University, Beijing 100084, China.
\\ [1.5cm]}

\Abstract{We evaluate the string theory disc amplitude of one Ramond-Ramond field $C^{(p-3)}$
and two Neveu-Schwarz $B$-fields in the presence of a single D$p$-brane in type II
string theory. From this amplitude we extract the
four-derivative (or equivalently order $\a'^2$) part of the
D$p$-brane action involving these fields.
We show that the new couplings are invariant under
R-R and NS-NS gauge transformations and compatible
with linear T-duality.}

\email{kbecker, guangyu, robbins@physics.tamu.edu}

\makepapertitle

\body


\tableofcontents

\newpage


\section{Introduction}

In this paper we continue the analysis of higher derivative contributions to the D$p$-brane action involving
one R-R potential and two NS-NS fields.
We will present the complete four-derivative action involving a R-R potential
of degree $p-3$, and two $B_{NS}$ fields. To do this we will compute world-sheet amplitudes with disc topology and
insertions of closed and open string vertex operators.

In ref. \cite{Becker:2010ij} we obtained part of the interactions. First we required that the D$p$-brane action
should be compatible with T-duality (see for example \cite{Bergshoeff:1996cy,Alvarez:1996up,Myers:1999ps}), which means that the dimensional reduction of a D$p$-brane should be related by
T-duality to the double dimensional reduction of a D$(p+1)$-brane. We used this requirement to predict some
four-derivative terms in the D$p$-brane Lagrangian. T-duality, however, does not determine the Lagrangian uniquely since
it can only be used in spaces with an isometry.
We verified the predictions from T-duality by computing scattering amplitudes for some choices of
polarization.
The predicted terms in the Lagrangian could easily be obtained from string amplitudes since
in the field theory limit only contact interactions on the brane contributed to these particular terms.
 The interactions predicted by
T-duality and the results obtained from the string theory amplitude in the limit of small momenta
did agree. However, the couplings were very special. In general,
a string amplitude with some vertex operator insertions can degenerate into many possible field theory diagrams. Most of these
diagrams are ``background noise'', by which we mean field theory diagrams which are constructed from known vertices either
in space-time or on the brane, and which need to be subtracted to isolate the field theory diagrams
which involve the new interactions. In general, this is a cumbersome procedure.
In this paper we have applied it to obtain the four-derivative
terms in the D$p$-brane effective action involving $C^{(p-3)}$ and two $B_{NS}$ fields.

\section{Overview and summary of results}

We start by summarizing our findings and will describe the details of our computations in the main part of the paper.
We wish to obtain the D$p$-brane action involving one R-R potential
$C^{(p-3)}$ and two $B_{NS}$ fields. To do this we will compute the
tree level string theory amplitude involving the vertex operators of
one R-R potential $C^{(p-3)}$ and
two $B_{NS}$ fields in the background of a D$p$-brane. The world-sheet has the topology
of a disc with insertions.
Schematically the amplitude is represented in fig.\ (1).
\begin{figure}[!htp]
\centerline{\includegraphics[scale=.7]{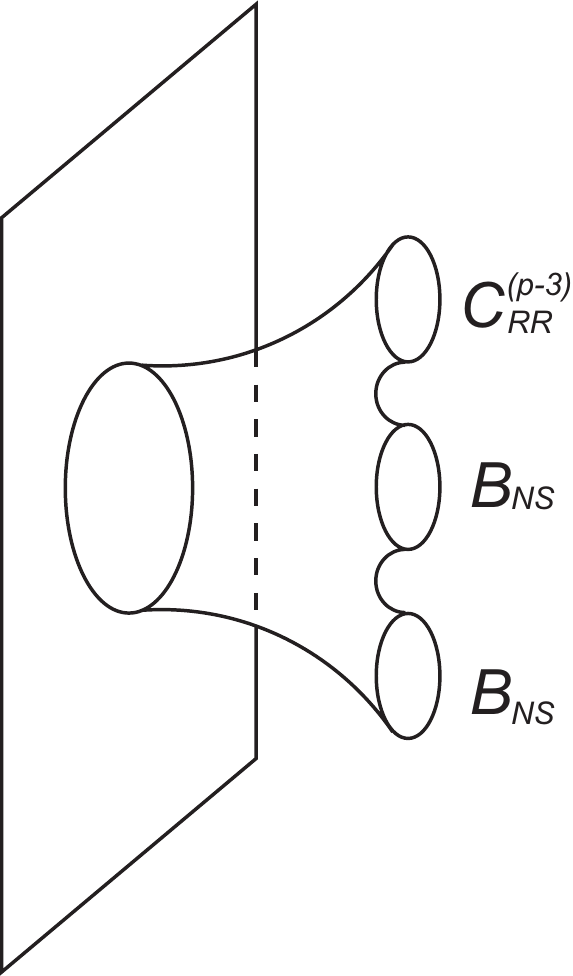}}
\label{fig:fig1}
\caption{String amplitude with closed string vertex operators in a D$p$-brane background.}
\end{figure}

Since the string amplitudes are invariant under
the parity transformation described in section 5.2 of ref. \cite{Becker:2011bw} it is easy to see
that string amplitudes involving $C^{(p-3)}$ are only non-vanishing if the two NS-NS fields
are both gravitons (or dilatons) or both $B_{NS}$ fields. This translates into the same statement for the D$p$-brane effective action
to all orders in $\a'$. This also generalizes to arbitrary R-R potentials in the following way. Amplitudes involving
$C^{(p-1+2k)}$ for $k$ odd are non-zero only if the two NS-NS fields have the same polarization, which means
both are symmetric or both are antisymmetric.
If $k$ is even the two NS-NS fields are required to have opposite polarizations.
If $k=-1$, which is the case considered here, the three-point amplitudes of $C^{(p-3)}$ and two gravitons
and the corresponding terms in the brane effective action
have
been found in refs. \cite{Bershadsky:1995qy,Green:1996dd,Craps:1998fn,Craps:1998tw,Morales:1998ux,Scrucca:1999uz,Stefanski:1998yx,Bachas:1999um}.
We will consider the case in which the NS-NS fields have generic anti-symmetric polarizations. We will compute
the string amplitudes and extract from them the D$p$-brane effective action to fourth order in derivatives.

We will obtain the string amplitudes in closed form only in an expansion in $\a'$ since only in this limit can we obtain
closed expressions for the complex integrals involved. These results are sufficient to extract
the four-derivative contributions to the D$p$-brane effective action. To obtain the effective action a careful
comparison with field theory amplitudes has to be performed.
For small momenta the string amplitude degenerates into six field theory diagrams displayed in fig.\ (2).
\begin{figure}[!htp]
\centerline{\includegraphics[scale=0.35]{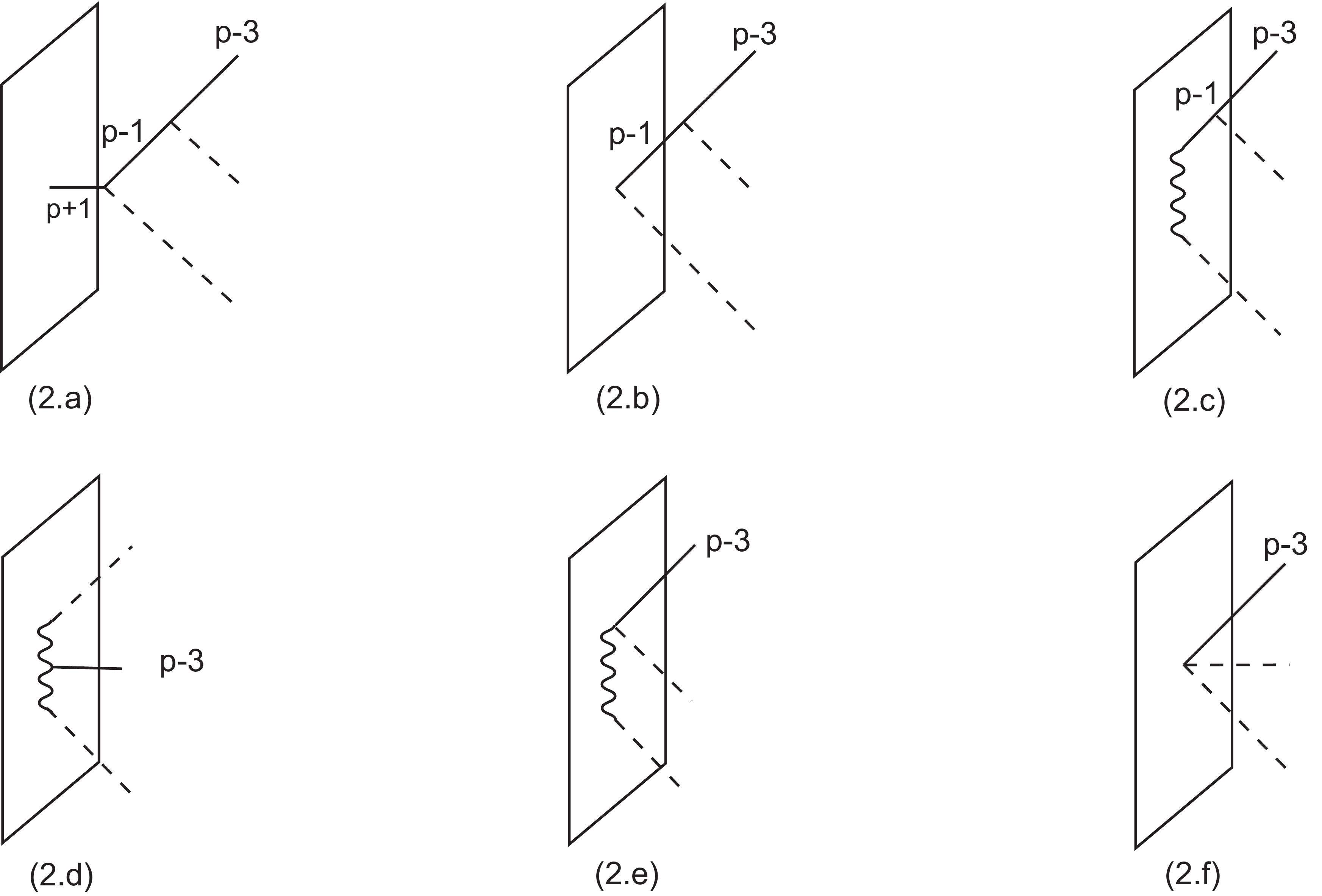}}
\label{fig:fig5}
\caption{Field theory diagrams arising from the string amplitude in fig.\ (1) in the limit of small momenta.
The dashed lines represent $B_{NS}$ fields, the wavy lines gauge fields on the D$p$-brane and
the solid lines are R-R fields of degree specified by the labels. }
\end{figure}
Diagram (2.f) represents the interaction of one R-R field and two NS-NS fields through a contact term on the
brane.
We wish to obtain this contact term to fourth order in derivatives. This encodes the corresponding
term in the D$p$-brane effective action. We will expand the string theory amplitude to quadratic order
in $\a'$ (fourth order in momenta) and subtract the result for diagrams (2.a)-(2.e).
This should give us the desired contact term.
To leading order in $\a'$ the diagrams (2.a)-(2.e) are, of course, known. However, these diagrams
themselves receive $\a'$ corrections arising from the corrections to contact terms on the D$p$-brane world-volume.
Specifically 3 contact terms receive corrections to order $\a'^2$. These are
displayed in fig.\ (3). These $\a'^2$ corrections are obtained by computing three
amplitudes. One two-point function involving a R-R potential $C^{(p-1)}$ and one $B_{NS}$ field, a
three-point amplitude involving 2 gauge fields on the brane and one R-R potential $C^{(p-3)}$ and
another three-point function involving one gauge field on the brane, a R-R potential $C^{(p-3)}$ and an $B_{NS}$ field.
We obtain the $\a'$ corrected contact terms by expanding to fourth order in momenta.
\begin{figure}[!htp]
\centerline{\includegraphics[scale=0.4]{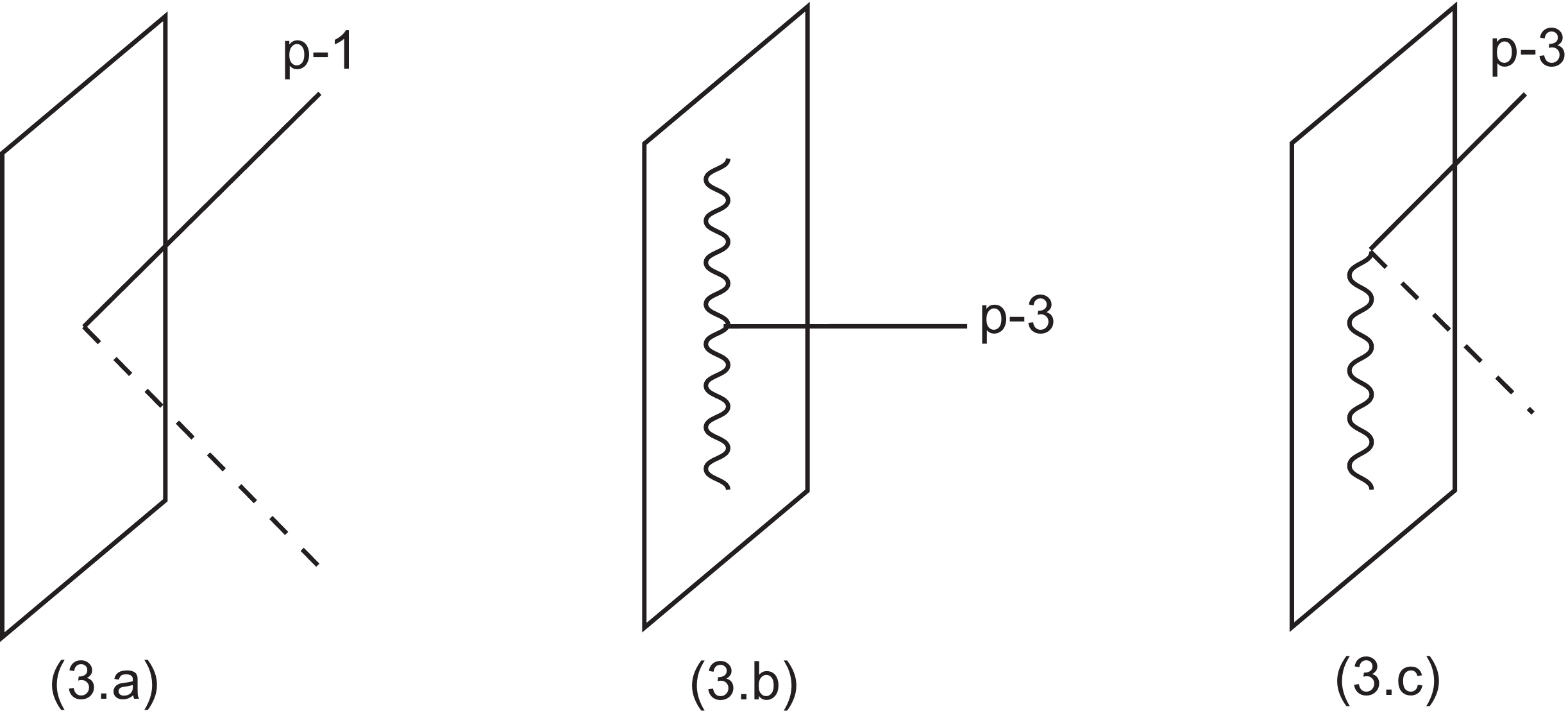}}
\label{fig:fig6}
\caption{Contact terms which are corrected to order $\a'^2$.}
\end{figure}
From here we extract the $\a'$ corrected result for diagrams (2.b), (2.d) and (2.e) (we will
show below that diagrams (2.a) and (2.c) do not receive corrections at order $\a'^2$).

The effective action is, of course, not unique. There are many effective actions which give rise to the same scattering
amplitudes. This ambiguity is related to the freedom in the choice of fields arising from field redefinitions.
Up to field redefinitions the Wess-Zumino part of the D$p$-brane effective action involving a R-R potential $C^{(p-3)}$, four-derivatives and two $B_{NS}$ fields is
\begin{equation}\label{action1}
{\mathcal L}={\mathcal L}^{(0)}+{\mathcal L}^{(4)} ,
\end{equation}
where
\begin{equation}
{\mathcal L}^{(0)}= {T_p\o 2^ 3 \G (p-2)} \e^{a_1 \cdots a_{p+1}}  {\tilde B}_{a_1a_2}{\tilde B}_{a_3a_4}C_{a_5\cdots a_{p+1}}^{(p-3)}
\end{equation}
and
\begin{equation}
\label{eq:FinalResult}
\begin{split}
{\mathcal L}^{(4)} = &
 {T_p I_0\o 4 \pi^2}\e^{a_1\cdots a_{p+1}}\bigg\{ \big( -\nabla^b {H_{a_1 b}}^i H_{a_2 a_3 i} - {1\o 2} \nabla^b {H_{a_1 a_2}}^i H_{a_3 b i } - {1\o 2} \nabla^i {H_{a_1 a_2 }}^b  H_{a_3 b i }\cr
&  +\nabla_b^{bc} \tilde B_{a_1 a_2} \tilde B_{a_3 c} + 2 \nabla^{bc} \tilde B_{a_1 a_2} \nabla_b \tilde B_{a_3 c} + \nabla_b^b {\tilde B_{a_1}}^{\hphantom{a_1}c}\nabla_c \tilde B_{a_2 a_3}   \big) F^{(p-2)}_{a_4 \cdots a_{p+1}} {1\over \G(p-1)} \cr
& + \left(\nabla^{bi} H_{a_1 a_2 b} - {1\o 2} \nabla^b_b {H_{a_1 a_2}}^i\right) \left[ {2\o \G(p) } \tilde F^{(p)}_{i a_3 \dots a_{p+1}} +{1\o \G(p-2)} \tilde B_{a_3 a_4} F^{(p-2)}_{i a_5 \dots a_{p+1}}\right]\bigg\} ,
\end{split}
\end{equation}
Here $\tilde B = B_{NS} + 2\a' F$, $F$ is the $U(1)$ gauge field, $\tilde F^{(n)} = d C^{(n-1)}  + H \wedge C^{(n-3)}$ is the R-R field strength and $H=dB_{NS}$. Throughout the paper we will use the convention that letters from the beginning of the Roman alphabet ($a$, $b$, {etc.}) indicate directions along the brane, while letters from the middle of the alphabet ($i$, $j$, {etc.}) indicate transverse directions.  Greek letters ($\m$, $\n$, {etc.}) run over all ten coordinates of the bulk space-time. Moreover, $\e$ is the volume form on the brane, $T_p$ the string tension and $I_0$ a constant.

The action (\ref{eq:FinalResult}) is the main result of this paper.  In the next section we will explain in detail the computation of the string amplitudes, their expansion
in powers of momenta and how to extract the D$p$-brane effective action.

\section{The details}

In this section we will describe the computation of the different string scattering amplitudes. We start by
presenting a formal definition of the $n$-point function and by proving that amplitudes will be independent of
the distribution of superghost charge as long as the total amount is $-2$, a property which will come very handy in concrete computations.

\subsection{General properties of string disc amplitudes}

Generalizing the construction of the two-point function described in ref. \cite{Becker:2011bw} we define
the $n$-point function on the disc by
\begin{multline}
\label{eq:GeneralNPoint}
\left\langle V^{(1)}(z_1,\bar{z}_1)V^{(2)}(z_2,\bar{z}_2)\left(\prod_{k=3}^n\int_{\CC}  d^2z_kU^{(k)}(z_k,\bar{z}_k)\right)\right.\\
\left.\times\left(b_0+\widetilde{b}_0\right)\int_{|w|>\operatorname{max}(1/|z_i|)}\frac{d^2w}{|w|^2}w^{-L_0}\bar{w}^{-\widetilde{L}_0}\left.\vphantom{\left(\prod_{k=3}^n\int d^2z_kU^{(k)}(z_k,\bar{z}_k)\right)}\right|B\right\rangle.
\end{multline}
Here we have introduced integrated vertex operators $U(z,\bar{z})$ which are related to physical state operators $V(z,\bar{z})$ by
\begin{equation}
U(z,\bar{z})=\left\{\widetilde{b}_{-1},\left[b_{-1},V(z,\bar{z})\right]\right\}.
\end{equation}
Saying that $V(z,\bar{z})$ is physical means that it is BRST-closed, has total left- plus right-ghost charge two, and has conformal weight zero on both the left and the right.  The second line of (\ref{eq:GeneralNPoint}) contains a boundary state~\cite{Callan:1987px,Kostelecky:1987px} and a propagator which expands it out until it hits the first insertion point (see fig.\ (4)).

If we pull the factor $w^{-L_0}\bar{w}^{-\tilde{L}_0}$ to the left of the correlator, we can rewrite this expression as
\begin{multline}
\left(\prod_{k=3}^n\int d^2z_k\right)\int_{|w|>\operatorname{max}(1/|z_i|)}d^2w|w|^{2n-6}\left\langle V^{(1)}(wz_1,\bar{w}\bar{z}_1)V^{(2)}(wz_2,\bar{w}\bar{z}_2)\vphantom{\left(\prod_{k=3}^nU^{(k)}(wz_k,\bar{w}\bar{z}_k)\right)}\right.\\
\left.\left.\times\left(\prod_{k=3}^nU^{(k)}(wz_k,\bar{w}\bar{z}_k)\right)\left(b_0+\widetilde{b}_0\right)\right|B\right\rangle\\
=\int_{|y_2|>1}\frac{d^2y_2}{|y_2|^2}\left\langle V^{(1)}(\infty,\infty)V^{(2)}(y_2,\bar{y}_2)\vphantom{\left(\prod_{k=3}^n\int_{|y_k|>1}d^2y_kU^{(k)}(y_k,\bar{y}_k)\right)}\right.\\
\left.\left.\times\left(\prod_{k=3}^n\int_{|y_k|>1}d^2y_kU^{(k)}(y_k,\bar{y}_k)\right)\left(b_0+\widetilde{b}_0\right)\right|B\right\rangle.
\end{multline}
In the second step we have taken advantage of the fact that the amplitude should be independent of $z_1$ to send $z_1$ to infinity, and then we have made changes of coordinate, $y_k=wz_k$.
\begin{figure}[!htp]
\centerline{\includegraphics[scale=0.38]{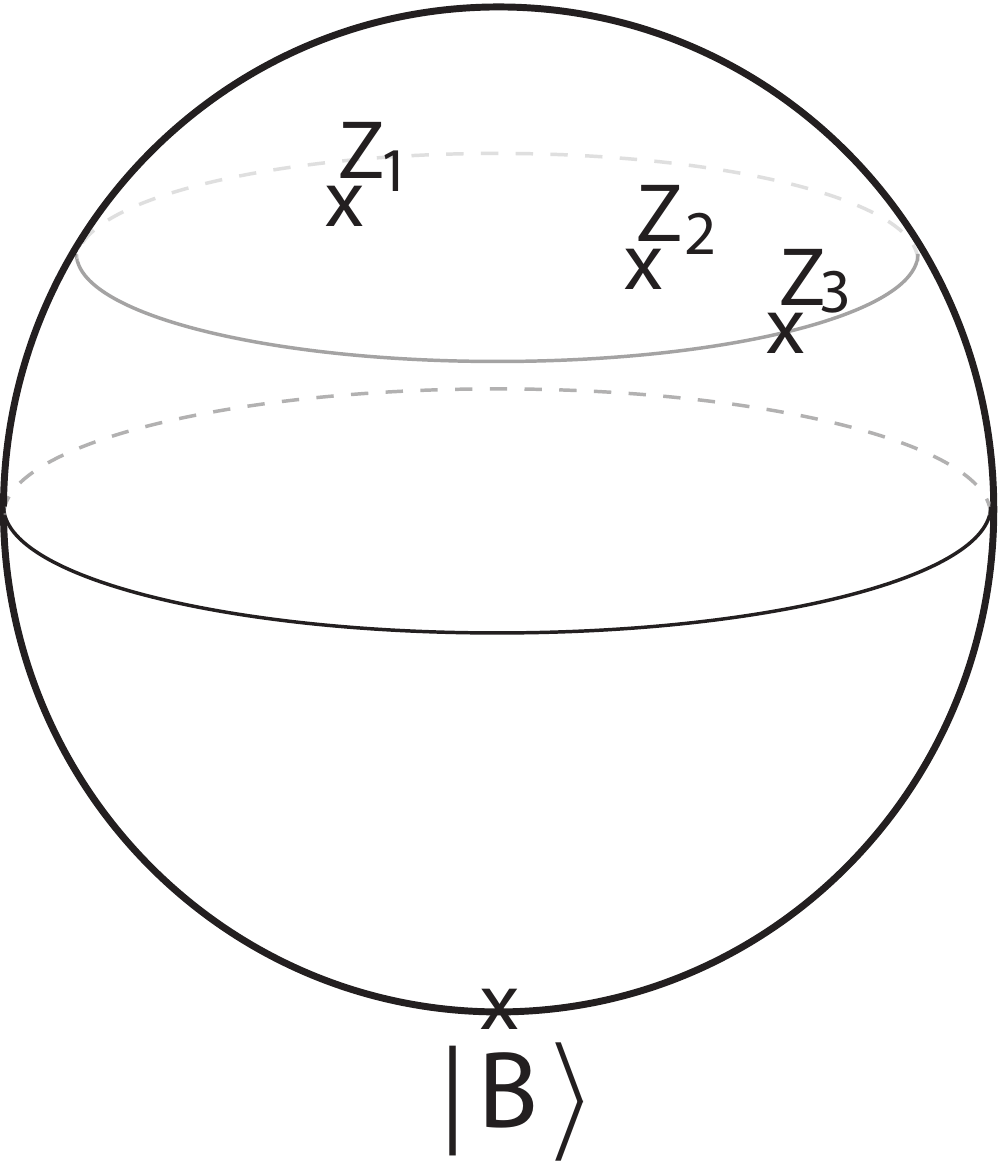}}
\label{fig:sp}
\caption{This figure illustrates a three-point function on the disc. A boundary state is created at one point on the sphere, in this case the south pole and transported using a propagator
to the location of the first insertion, in this case $z_3$. }
\end{figure}

It is easy to check that (up to total derivatives)
$U(z,\bar z)$ is BRST closed if $V(z,\bar z)$ is and if $V(z, \bar z)$ is BRST exact so is
$U(z, \bar z)$ (again, up to total derivatives). These total derivatives give rise to boundary
terms which vanish for an entire range of momenta and therefore analytic continuation guarantees that
the boundary terms vanish everywhere. Therefore as long as the $V(z,\bar z)$ vertex operators are
BRST closed, BRST trivial states will decouple from $n$-point functions.

Note that the picture changing operator $X_0$ does not commute with $b_{-1}$. Rather,
\begin{equation}
[X_0, b_{-1}] = -2 \xi_{-1},
\end{equation}
and therefore the independence of the $n$-point functions of the distribution of picture charge requires
a careful treatment. The simplest way to show the picture independence of $n$-point functions is
to check that $[X_0, U(z,\bar z)]$ is zero up to BRST trivial pieces and total derivatives, which in turn vanish
using analytic continuation. Indeed, as can be easily verified
\begin{equation}
\begin{split}
&[X_0, U(z,\bar z)]  = \oint_{|w-z|=\e}\frac{dw}{2\pi iw}X(w)U(z,\bar z)  = \cr
& \oint_{|w-z|=\e}\frac{dw}{2\pi iw}\lp X(w_0)+\int_{w_0}^wdu \pa X(u)\rp U(z,\bar{z}) = \cr
& \{ Q, 2 \oint { d w \o 2 \pi i w} \x(w) U(z, \bar z)\} - 2 {\pa \over \pa \bar z} \left( \oint { dw \over 2 \pi i w } \{ b_{-1}, \xi(w) V(z, \bar z) \} \right) \cr
& + 2 {\pa \over \pa z} \left( \oint { dw \over 2 \pi i w} \{ \tilde b_{-1} , \xi(w) V(z, \bar z) \}\right).
\end{split}
\end{equation}
Since the picture changing operator $X_0$ commutes with the two types of vertex operators, integrated and non-integrated
ones, $n$-point functions will be independent of the distribution of picture charge. This is a very useful property since it is a way of
checking our results. The string amplitudes we will compute are not manifestly picture independent and different contributions have
to combine in a non-trivial way to give rise to a picture independent result.

\subsection{Amplitudes involving closed string vertex operators}

\subsubsection{One R-R field and one $B_{NS}$ field}

The two-point of one R-R field and one $B_{NS}$ field has been computed before. The original references are \cite{Gubser:1996wt,Garousi:1996ad,Hashimoto:1996kf,Hashimoto:1996bf,Garousi:2010ki}
or using the notation and conventions of this paper in ref. \cite{Becker:2011bw}. We will
label any disc string amplitude by ${\cal A}$ and any field theory amplitude by ${\bf A}$  with indices specifying the vertex operator insertions.
In a form convenient for our purposes the 2-point function of $C^{(p-1)}$ and $B_{\rm NS}$ is
\begin{equation}
\begin{split} \label{eq:stringbc}
{\cal A}_{CB}  = & {T_p\o 2 \G(p)}\begin{pmatrix} p_2 Dp_2 +{(p_1+p_2)^2 \o 2}\\ p_2 Dp_2 \end{pmatrix}^{-1}
\e^{a_1\cdots {a_{p+1}}}\bigg\{ C^{(p-1)}_{a_1\cdots a_{p-1}} \bigg[ {2(p_2D\e)_{a_p} (p_2)_{a_{p+1}} \o (p_1+ p_2)^2} \cr
& + {2(p_2D\e)_{a_p}(p_2)_{a_{p+1}}\o p_2Dp_2}   -{2(p_1\e)_{a_p}(p_2)_{a_{p+1}}\o (p_1+ p_2)^2} + \left(1+{p_2Dp_2-p_1\cdot p_2\o ( p_1+ p_2)^2}\right)\e_{a_p{a_{p+1}}}\bigg] \cr &
-{(p-1)\o ( p_1+ p_2)^2}C^{(p-1)} _{a_1\cdots a_{p-2}\mu}(p_2)_{a_{p-1}}(Dp_2)^\mu\e_{a_p{a_{p+1}}} \bigg\}.
\end{split}
\end{equation}
We have introduced the matrix $D^\m_{\hphantom{\m}\n}$ which is diagonal with entries $+1$ in the directions along the brane and $-1$ in the directions normal to the brane.  On-shell this agrees with the result quoted in  ref. \cite{Becker:2011bw} as can be easily verified. Here $(\dots)$ denotes the binomial coefficient, and $\mu$ represents a space-time index that is summed over both tangent and normal directions.

\subsubsection{One R-R field and two $B_{NS}$ fields}

The disc amplitude in fig.\ 1 is
\begin{equation}
{\cal A}_{CBB}=\langle V^{(-1/ 2,-1/ 2)}_{C}(p_1)V^{(-1,0)}_{B}(\e_2, p_2) V^{(0,0)}_B(\e_3, p_3)|B\rangle.
\end{equation}
We will take the R-R vertex operator in the $(-1/2,-1/2)$ picture.
As a result the $B_{NS}$ fields have to
be in different superghost pictures which we take to be $(0,0)$ and $(-1,0)$. So the amplitude is not manifestly
invariant under the interchange of the two polarization tensors. But since the amplitudes are picture independent the
result should be symmetric which will be a non-trivial check of our results.

We will separate the amplitude into various pieces according to their index structure and use the notation
\begin{equation}\label{eq:stringamplitude}
{\cal A}_{CBB}=
\sum_{i=1}^5 {\cal A}_{CBB}^{(i)} .
\end{equation}
In the following we will quote our results for ${\cal A}_{CBB}^{(i)}$
\begin{enumerate}
\item[1)] \underline{$(\e_2\cdot p)(\e_3\cdot p)$ and $(\e_2\cdot \e_3)$ terms}

The sum of the terms proportional to either $(\e_2\cdot p)(\e_3\cdot p)$
or $(\e_2\cdot \e_3)$ for arbitrary polarization tensors $\e_2$ and $\e_3$ is
\begin{equation}
\begin{split}\label{A1}
{\cal A}_{CBB}^{(1)}=
&{i\o2\sqrt{2}}{1\o (p-2)!}\e^{a_1\cdots a_{p+1}} (p_2)_{a_1}(p_3)_{a_2} C_{a_5\cdots a_{p+1}}^{(p-3)}\bigg[(p_2p_3)(\e_2D\e_3){}_{a_3a_4}I_0\cr
& -(p_2Dp_3)(\e_2\e_3){}_{a_3a_4}I_0+(p_2D\e_2)_{a_3}(p_3D\e_3)_{a_4}I_3 -(p_2D\e_2)_{a_3}(p_2D\e_3 )_{a_4}I_7\cr
& +(p_3D\e_2 )_{a_3}(p_2\cdot\e_3 )_{a_4}I_8-(p_2D\e_2 )_{a_3}(p_2\cdot\e_3 )_{a_4}I_6-(p_2D\e_2 )_{a_3}(p_1N\e_3 )_{a_4}I_4\cr
& -(p_3D\e_2 )_{a_3}(p_1N\e_3 )_{a_4}I_5+(p_3\cdot\e_2)_{a_3}(p_1N\e_3 )_{a_4}I_9  +(p_1N\e_2 )_{a_3}(p_1N\e_3 )_{a_4}I_{10} \bigg] \cr
& +\left(2\leftrightarrow 3\right)  ,
\end{split}
\end{equation}

Here $I_{n}$ are integrals whose definition
and whose approximate values in the region of small momenta can be found in the appendix A. Moreover
\begin{equation}
N_{\m\n} = {1\o 2} \left( \eta_{\m\n} - D_{\m\n}\right) ,
\end{equation}
and by $(2 \leftrightarrow 3)$ we mean the same expression but interchanging $p_2 \leftrightarrow p_3$ and $\e_2 \leftrightarrow \e_3$.

\item[2)] \underline{$(p\cdot\e \cdot p)(\e)$ term}

The sum of terms of index structure $(p\cdot\e \cdot p)(\e)$ is
\begin{equation}
\label{A2}
\begin{split}
{\cal A}_{CBB}^{(2)} =& {i\o4\sqrt{2}}{1\o (p-2)!} \e^{a_1\cdots a_{p+1}}(p_2)_{a_1}(p_3)_{a_2}(\e_2)_{a_3a_4} C_{a_5\cdots a_{p+1}}^{(p-3)} \bigg[(p_2\e_3Dp_3)I'_6 + (p_2D\e_3Dp_3)I'_7\cr
& +(p_2D\e_3Np_1)I_5 -(p_2\cdot\e_3Np_1)I_9+(p_2\e_3Dp_2)(I_8-2I_0)\bigg] + \left( 2\leftrightarrow 3\right),
\end{split}
\end{equation}
where $I_n'=I_n(p_2\leftrightarrow p_3)$.

\item[3)] \underline{$(\e \cdot p)(\e)$ terms}

There are two terms with the above quoted index structure, one with all
indices of $C^{(p-3)}$ along the brane and another one in which
one of the indices is transverse to the brane. The first one is given by
\begin{equation}\label{A3}
\begin{split}
{\cal A}_{CBB}^{(3)}=
&{i\o 8\sqrt{2}}{1\o (p-2)!} \e^{a_1\cdots a_{p+1}}(p_2)_{a_1}(\e_2)_{a_2a_3}C_{a_5\cdots a_{p+1}}^{(p-3)}\bigg[(p_2\cdot p_3)(p_2\cdot\e_3)_{a_4}I_2\cr
& -(p_2Dp_3)(p_2D\e_3)_{a_4}I_1+2(p_2Dp_3) (p_2\cdot\e_3)_{a_4}I_0 -2(p_2\cdot p_3)(p_2D\e_3)_{a_4}I_0\cr
& -(p_2Dp_3) (p_2\cdot\e_3)_{a_4}I_8 +(p_2\cdot p_3)(p_2D\e_3)_{a_4}I_8+2(p_2D p_3)(p_3D\e_3)_{a_4}I'_7 \cr
& + 2(p_2\cdot p_3)(p_3D\e_3)_{a_4}I'_6-(p_3D p_3)(p_2\cdot \e_3)_{a_4}I'_6-(p_3D p_3)(p_2D \e_3)_{a_4}I'_7\cr
& -2(p_2\cdot p_3)(p_1N\e_3)_{a_4}I_9+2(p_2Dp_3)(p_1N\e_3)_{a_4}I_5\bigg]\cr
& +{i\o 8\sqrt{2}}{1\o (p-2)!} \e^{a_1\cdots a_{p+1}}(p_2)_{a_1}(\e_3)_{a_3a_4}C_{a_5\cdots a_{p+1}}^{(p-3)}\bigg[(p_2\cdot p_3)(p_3\cdot \e_2)_{a_2} I_2\cr
&-(p_2D p_3)(p_3D \e_2)_{a_2}I_1+2(p_2D p_3)(p_3\cdot \e_2)_{a_2}I_0 -2(p_2\cdot p_3)(p_3D \e_2)_{a_2}I_0\cr
&-(p_3Dp_3)(p_1N\e_2)_{a_2}I_4'-(p_3Dp_3)(p_3\cdot \e_2)_{a_2}I_6' -(p_3Dp_3)(p_3D \e_2)_{a_2}I_7'\cr
&+(p_2\cdot p_3)(p_3D \e_2)_{a_2}I_8 -(p_2D p_3)(p_3\cdot \e_2)_{a_2}I_8+2(p_3D p_3)(p_2D \e_2)_{a_2}I_3\bigg]\cr
&+\left( 2 \leftrightarrow 3\right).
\end{split}
\end{equation}
This result is manifestly symmetric under the interchange of the two $B_{NS}$ fields; to write it this way we have made use of certain relations between the $I_n$ which follow from the definitions and expansions in the appendix.
\item[4)] \underline{$(\e_2)(\e_3)$ terms}

Terms with the above index structure are
\begin{equation}\label{A4}
\begin{split}
{\cal A}_{CBB}^{(4)} =&{i\o 4\sqrt{2}} \e^{a_1\cdots a_{p+1}}\Bigg\{ {1\o 4}  {1\o (p-2)!}C^{(p-3)}_{a_5\cdots a_{p+1}} (\e_2)_{a_1a_2}(\e_3)_{a_3a_4} \bigg[(p_2Dp_3)^2I_1-(p_2\cdot p_3)^2I_2\cr
& -(p_2Dp_2)(p_3Dp_3)I_3 \bigg] +{1 \o 4}{p-3\o (p-2)!}  C^{(p-3)}_{ia_6\cdots a_{p+1}}(\e_2)_{a_1a_2}(\e_3)_{a_3a_4}\times \cr
& \bigg[2(p_3)_{a_5} p_2^i(p_2Dp_3) I_5+2(p_3)_{a_5} p_2^i(p_2\cdot p_3)I_9-(p_2)_{a_5} p_2^i(p_3Dp_3)I_4'\cr
& +2 (p_2)_{a_5} p_3^i (p_2Dp_3)I_5-2(p_2)_{a_5} p_3^i(p_2\cdot p_3)I_9-(p_3)_{a_5} p_3^i(p_2Dp_2)I_4\bigg]\cr
&+{(p-3)(p-4)\o (p-2)!} C^{(p-3)}_{ija_7\cdots a_{p+1}}(\e_2)_{a_3a_4} (\e_3)_{a_5a_6} (p_2)_{a_1}(p_3)_{a_2} p_3^ip_2^jI_{10}\Bigg\},
\end{split}
\end{equation}
where we have again used certain relations among the $I_n$ to write ${\cal A}_{CBB}^{(4)}$ in a way which is manifestly symmetric under exchange of the two NS-NS fields.

\item[5)] \underline{other $C^{(p-3)}_{ia_6\cdots a_{p+1}}$ terms}

For the case that one of the indices of the R-R potential is transverse to the brane the amplitude
contribution is
\begin{equation}\label{A5}
\begin{split}
{\cal A}_{CBB}^{(5)} =&{i\o 4\sqrt{2}}{p-3\o (p-2)!} \e^{a_1\cdots a_{p+1}} C_{ia_6\cdots a_{p+1}}(p_2)_{a_1}(p_3)_{a_2}(\e_2)_{a_3a_4}\bigg[p_3^i(p_2\cdot \e_3)_{a_5} I_9\cr
& -p_3^i(p_2D \e_3)_{a_5} I_5-2p_2^i(p_1N\e_3)_{a_5} I_{10}+p_2^i(p_2\cdot \e_3)_{a_5} I_9+p_2^i(p_2D\e_3)_{a_5} I_5\cr
& + p_2^i(p_3D\e_3)_{a_5}I_4'\bigg]+\left( 2 \leftrightarrow 3\right).
\end{split}
\end{equation}

\end{enumerate}

\subsection{Amplitudes involving closed string vertex operators and gauge fields}

\subsubsection{One R-R field and two gauge fields}

Let the string amplitude for a D$p$-brane
absorbing one R-R field and emitting two open string gauge fields
be
\begin{equation}
{\cal A}_{CAA}=\langle V^{(-{1/ 2},-{1/ 2})}(p_1)V^{-1}(p_2,
\zeta_2)V^{0}(p_3, \zeta_3)|B\rangle,
\end{equation}
where following ref. \cite{Hashimoto:1996bf}, the open string vertex operators are
\begin{equation}
\begin{split}
V^{-1}(p_2, \zeta_2)&=(\zeta_2)_a\int dxe^{-\phi}\psi^a e^{2ip_2\cdot X}(x)\cr
V^{0}(p_3, \zeta_3) & = (\zeta_3)_a \int dx(\pa X^a-2ip_3\cdot \psi \psi^a) e^{2ip_3\cdot X}(x).
\end{split}
\end{equation}
Here the momenta are constrained to be parallel to the brane and therefore $p_2\cdot p_3=p_2^ap_3^a$
and the vertex operators are integrated over the world-sheet boundary (a circle). The result for this
amplitude is
\begin{equation}
\begin{split}
{\cal A}_{CAA} = & \begin{pmatrix} 4 p_2 \cdot p_3 \\ 2 p_2 \cdot p_3 \end{pmatrix}
\e^{a_1\cdots a_{p+1}} F_{a_1a_2}F_{a_3a_4} C^{(p-3)}_{a_5\cdots a_{p+1}} \cr
= &    \bigg[1+{2\pi^2\o 3}(p_2\cdot
p_3)^2\bigg]\e^{a_1\cdots a_{p+1}}
F_{a_1a_2}F_{a_3a_4} C^{(p-3)}_{a_5\cdots a_{p+1}}+\dots .
\end{split}
\end{equation}
Here
\be
F_{ab}=ip_a\zeta_b-ip_b\zeta_a,
\ee
is the field strength for the gauge field with polarization $\zeta_a$.


\subsubsection{One R-R field, one $B_{NS}$ field and one gauge field}

Let the disc amplitude with insertions of one R-R vertex operator, one $B_{NS}$ field, and
one open string vertex operators be
\begin{equation}
{\mathcal A}_{CAB}=\langle V^{(-{1/ 2},-{1/
2})}(p_1)V^{-1}(p_2,\zeta)V^{(0,0)}(p_3,\e)|B\rangle.
\end{equation}
The result is
\begin{equation}
\begin{split}
{\mathcal A}_{CAB} =&{i\o 2^{7/2}}{1\o (p-3)!}\e ^{a_1\cdots a_{p+1}}C_{a_5\cdots a_{p+1}}\bigg[\e_{a_1a_2}(p_3)_{a_3}\zeta_{a_4}\left({p_2\cdot p_3\o p_1\cdot p_3}Q-{2\pi^2\o 3}(p_2\cdot p_3)^2\right)\cr
&-\e_{a_1a_2}(p_3)_{a_3}(p_2)_{a_4}(p_3\cdot \zeta)\left({Q\o p_1\cdot p_3}-{2\pi^2\o 3}p_2\cdot p_3\right)\cr
&+F_{a_1a_2} (p_3)_{a_4}(p_2\cdot\e)_{a_3}{2\pi^2\o 3}(p_2\cdot p_3)\cr
&-F_{a_1a_2}(p_3)_{a_4}(p_3D\e)_{a_3}\left({2\o p_3Dp_3}Q+{4\pi^2\o 3}{(p_2\cdot p_3)^2\o p_3Dp_3}+{Q\o 2p_1\cdot p_3}\right)\cr
&+F_{a_1a_2}(p_3)_{a_4}(p_1\cdot\e)_{a_3}{Q\o p_1\cdot p_3} \cr
&-F_{a_1a_2}\e_{a_3a_4}\left({1\o 2}Q+{1\o 4}{p_3Dp_3\o p_1\cdot p_3}Q+{\pi^2\o 3}(p_2\cdot p_3)^2\right)\bigg]\cr
&-{i\o 2^{9/2}}{1\o (p-4)!}\e ^{a_1\cdots a_{p+1}}F_{a_1a_2}\e_{a_3a_4}(p_3)_{a_5}p_3^i C_{i\,a_6\cdots a_{p+1}}{Q\o p_1\cdot p_3},
\end{split}
\end{equation}
and
\begin{equation}
Q=\begin{pmatrix} p_1 \cdot p_3 + p_3 Dp_3 \\ p_1 \cdot p_3 \end{pmatrix}^{-1}
= 1-{\pi^2\o 6}(p_1\cdot p_3)(p_3Dp_3)+\dots.
\end{equation}

\subsection{Derivative corrections to field theory vertices localized on the D$p$-brane.}

The string amplitudes encode all field theory vertices. In this section we describe how to obtain the effective Lagrangian on the D$p$-brane
world-volume from the expansion about small momenta of the string theory amplitudes.

\subsubsection{$C^{(p-1)}$, $B$ contact term}
\label{subsubsec:CB}

Next we will start with the 2-point function of a R-R potential $C^{(p-1)}$ and an NS-NS
B field and use it to derive four-derivative corrections to
the D-brane action which are quadratic in the fields.  
In the field theory approximation the string amplitude gives rise to 3 diagrams displayed in fig.\ (5).
\begin{figure}[!htp]
\centerline{\includegraphics[scale=0.45]{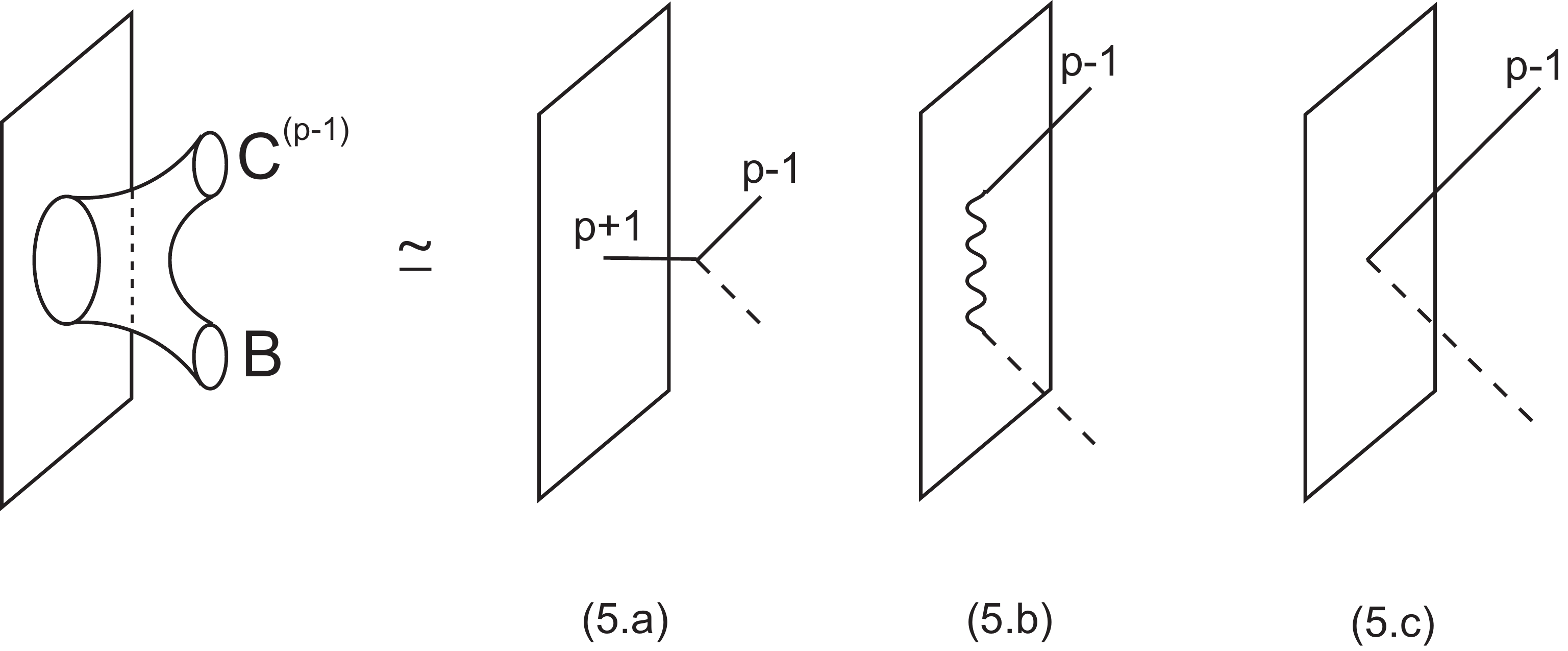}}
\label{fig:2pt}
\caption{Field theory diagrams arising in the small momentum limit
of a string scattering amplitude involving $C^{(p-1)}$ and an $B_{NS}$ field in a D$p$-brane background. }
\end{figure}
In this section we will obtain the four-derivative correction to the vertex in fig.\ (5.c) by expanding eqn. (\ref{eq:stringbc}) in powers of $\a'$.
The leading term is the field theory result which we need to substract.

Note that the amplitudes (5.a) and (5.b) do not receive corrections to order $\a'^2$ which is the order we are interested in. This allows us to directly
obtain the $\a'^2$ corrections to the vertex in fig.\ (5.c).
The field theory result for figures (5.a) and (5.b) can be rewritten as\footnote{In  ref. \cite{Becker:2011bw} we derived all relevant field theory diagrams in our
conventions. We refer the reader to ref. \cite{Becker:2011bw} for more details.   }
\begin{equation}
\begin{split}
{\bf A}^{(a)}_{BC}= & - {T_p\o 4 \G(p) }\e^{a_1\cdots a_{p+1}}\bigg\{{2 (p-1)\o (p_1+p_2)^2}C^{(p-1)} _{a_1\cdots a_{p-2}i}\e_{a_{p-1}a_p}(p_1)_{a_{p+1}}p_2^i\cr
& + C^{(p-1)}_{a_1\cdots a_{p-1}}\bigg[ \left(1-{p_2Dp_2\o (p_1+p_2)^2}\right)\e_{a_pa_{p+1}} +{8\e_{a_pb} (p_1)_{a_{p+1}}p_1^b\o (p_1+p_2)^2}+{4\e_{a_pi} (p_1)_{a_{p+1}}p_1^i\o (p_1+p_2)^2}\bigg]\bigg\}
\end{split}
\end{equation}
and
\begin{equation}
{\bf A}^{(b)}_{CB}={T_p\o \G(p)}{2\o p_2Dp_2}\e^{a_1\cdots
a_{p+1}}C^{(p-1)}_{a_1\cdots a_{p-1}}\e_{ba_p} p_{1\,a_{p+1}}p_1^b
\end{equation}
Here and in the following we will label the field theory amplitudes by $\bf A$ with indices specifying the fields involved.

After subtracting the result for diagrams (5.a)
and (5.b) we obtain the result for the diagram (5.c), which can be derived from the following effective action
\begin{equation}
{\mathcal L}_{CB}   =   {\mathcal L}_{CB}^{(0)}  +{\mathcal L}_{CB}^{(4)},
\end{equation}
with
\begin{equation}\label{eq:ActionBC}
\begin{split}
{\mathcal L}_{CB}^{(0)}  = & {T_p\o 2 \G(p)}
\e^{a_1  \cdots a_{p+1}}  B_{a_1 a_2}C^{(p-1)}_{a_3\cdots a_{p+1}} \cr
{\mathcal L}_{CB}^{(4)}  =  {T_p\o \G(p)} {I_0 \over 4 \pi^2}
\e^{a_1  \cdots a_{p+1}}
 \Big[ &
- \nabla^a{}_aH_{a_1 a_2i}\nabla^i C^{(p-1)}_{a_3 \cdots a_{p+ 1}} +
2 \nabla^a{}_iH_{a_1 a_2a}\nabla^iC^{(p-1)}_{a_3\cdots a_{p+1}}
\cr  & +  {(p-1) \o 3 }\nabla^{ia}{}_aH_{a_1 a_2 a_3}C^{(p-1)}_{i\, a_4\cdots a_{p+1}}+\nabla^a H_{a_1 a_2 a} \nabla^\m _\m C^{(p-1)}_{a_3 \dots a_{p+1}} \Big].
\end{split}
\end{equation}
to order $\a'^2$. This effective action encodes the correction to the vertex in fig.\ (3.a).  It can be checked that this result agrees with \cite{Garousi:2010ki} up to terms which vanish on-shell.

The above effective action is obtained from on-shell amplitudes. As such it is ambiguous. We will describe some ambiguities with concrete examples.
Consider the one-point function of a R-R vertex operator in a D$p$-brane background,
which is exact in the derivative expansion. From
here we conclude that the D-brane action involving one R-R field,
$ S_{Dp}\sim \int C_{p+1}$ can only receive corrections
which vanish on-shell. This means that we could have added an
interaction of the form
\begin{equation} \label{bi}
\d {\bf A}_{Dp}  \sim  \a' \int \pa_\mu \pa^\m C_{p+1},
\end{equation}
for example, without changing
the 1-point function since this expression vanishes on-shell and string amplitudes involve on-shell vertex operators. However,
if the effective D$p$-brane action gets such a contribution, in principle higher
point functions could be modified since for general tree diagrams, a propagator
connecting a R-R field to the brane will involve off-shell momenta. So for example,
diagram (5.a) changes by
\begin{equation} \label{bii}
\d {\bf A}^{(a)}_{CB} \sim (p_1 + p_2)^2 {\bf A}^{(a)}_{CB}.
\end{equation}
The factor $(p_1+p_2)^2$ on the right hand side of eqn. \C{bii} cancels all such factors in the denominator of ${\bf A}^{(a)}_{CB}$. When extracting the result for the vertex in fig.\ (5.c) we will then obtain an expression
which is also shifted and which is very easy to work out (we do not need the details here). That
the correction to the one-point function vanishes on-shell guarantees that this shift is a contact term
representable as a contribution to the vertex in fig.\ (5.c). However, by construction
$ \d {\bf A}^{(a)}_{CB} + \d {\bf A}^{(c)}_{CB}=0,$
and as a result any shift in $\d {\bf A}^{(a)}_{CB} $ will be compensated by a shift in $\d {\bf A}^{(c)}_{CB}$ in such
a way that the two-point function of $B_{\rm NS-NS}$ and $C^{(p-1)}$ is left unchanged.  We conclude that any corrections
to the one-point function of a R-R field of the form \C{bi} do not change the one nor two-point functions of on-shell states.

\subsubsection{$C^{(p-3)}$, $A$, $A$ contact term.}

For small momenta the amplitude ${\cal A}_{CAA}$ can be obtained from the following effective action
\begin{equation} \label{actioncaa}
{\mathcal L}_{CAA}={T_p(2\a')^2
\o 2^3\G(p-2)}
\e^{a_1\cdots a_{p+1}}\left( F_{a_1a_2}
F_{a_3a_4}+{2\pi^2\o 3}\nabla^a{}_b F_{a_1a_2}\nabla^b{}_a
F_{a_3a_4}\right)C_{a_5\cdots a_{p+1}}^{(p-3)} ,
\end{equation}
again up to terms that vanish on-shell. The subleading term has been obtained before in ref. \cite{Wyllard:2000qe}, \cite{Wyllard:2001ye}.
This effective action includes the corrections to the vertex in fig.\ (3.b).

Note that we have left an explicit factor of $(2\a')$ in for each factor of $F_{ab}$, while in most of the paper we have set $\a'=2$.  Our main purpose for this is simply book-keeping; it serves as a reminder that we should not count the derivative used in constructing $F$ in our derivative expansion.  In other words, we treat the combination $2\a'F$ as having weight zero, consistent with the fact that we will be using it to build the gauge invariant objects $\tilde B_{ab}=B_{ab}+2\a'F_{ab}$.

\subsubsection{$C^{(p-3)}$, $A$, $B$ contact term.}

In the field theory limit the string amplitude ${\cal A}_{CAB}$ gives rise to the diagrams represented in fig.\ (6).
\begin{figure}[!htb]
\label{sugracab}
\begin{center}
\scalebox{.4}{\includegraphics{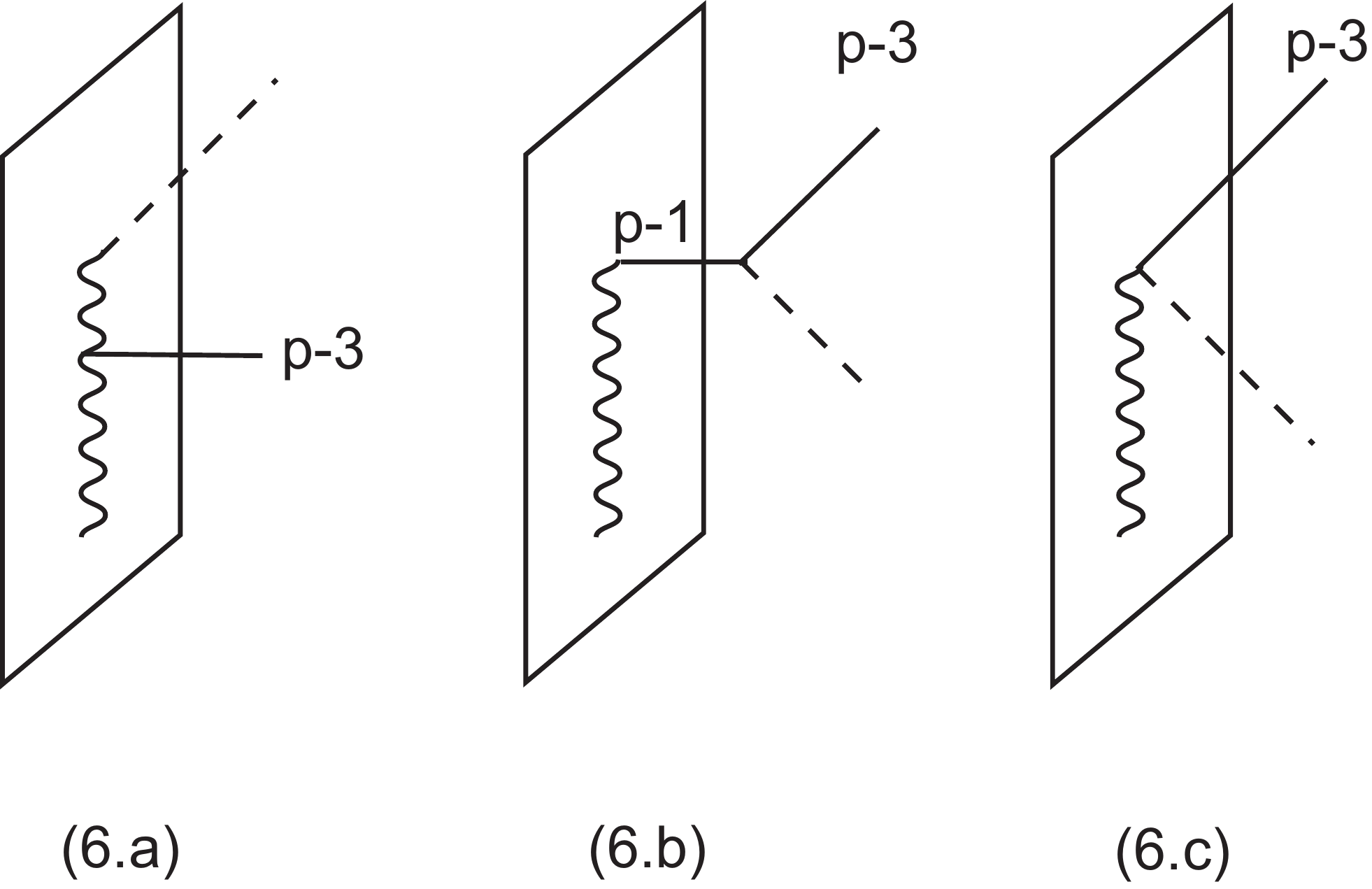}} 
\caption{ Field theory diagrams arising in the small momentum limit
of a string scattering amplitude involving $C^{(p-3)}$, a gauge field and an $B_{NS}$ field.}
\end{center}
\end{figure}
The amplitudes for the diagrams (6.a) and (6.b) are
\begin{equation}
{\bf A} ^{(a)}_{CAB}=-{i\o 2^{5/2}}{1\o (p-3)!}\e^{a_1 \cdots a_{p+1}}{1\over p_3 Dp_3}
 F_{a_1a_2}(p_3D\e)_{a_3}(p_3)_{a_4} C_{a_5\cdots a_{p+1}}^{(p-3)}
\left[ 1 +{2\pi^2\o 3}(p_2\cdot p_3)^2\right],
\end{equation}
and
\begin{eqnarray}
{\bf A}^{(b)}_{CAB} &=&{i\o 2^{7/2}}{1\o
(p-3)!}\e^{a _1\cdots a_{p+1}}
{1\over p_1 \cdot p_3} \bigg[-{1\o 2}F_{a _1a _2} (p_3)_{a _4}(p_3D\e)_{a _3} +F_{a _1a _2}(p_3)_{a _4}(p_1\cdot\e)_{a _3} \cr &&
+ p_2\cdot p_3 \e_{a _1a _2}(p_3)_{a _3}\zeta_{a _4} -\e_{a _1a _2}(p_3)_{a _3}(p_2)_{a _4}(p_3\cdot \zeta)-
  {1\o 4} p_3Dp_3  F_{a _1a _2}\e_{a _3a _4}\bigg]C_{a_5\cdots a_{p+1}}^{(p-3)} \cr
&&-{i\o 2^{9/2}}{1\o (p-4)!}{1\o p_1\cdot p_3} \e^{a _1\cdots a_{p+1}}F_{a _1a _2}\e_{a _3a _4}(p_3)_{a _5}p_3^i C_{i\,a_6\cdots a_{p+1}}^{(p-3)}\cr
&&+{i\o 2^{9/2}}{1\o (p-3)!}\e^{a _1\cdots a_{p+1}}F_{a _1a _2}\e_{a _3a _4}C_{a_5\cdots a_{p+1}}^{(p-3)} .
\end{eqnarray}
After subtracting the field theory amplitudes ${\bf A}^{(a)}_{CAB}$ and
${\bf A}^{(b)}_{CAB}$ (figs. (6.a) and (6.b)) from the string amplitude in the limit of small momenta, we obtain the
field theory amplitude ${\bf A}^{(c)}_{CAB}$ (fig.\ (6.c)),
which can be obtained from the following action
\begin{equation}
{\mathcal L}_{CAB}={\mathcal L}_{CAB}^{(0)} +{\mathcal L}_{CAB}^{(4)},
\end{equation}
where
\begin{equation}
{\mathcal L}_{CAB} ^{(0)}= {T_p\o 2^2 \G(p-2)}\e^{a _1\cdots a_{p+1}}B_{a _1a _2}(2\a'
F)_{a _3a _4}C_{a_5\cdots a_{p+1}},
\end{equation}

\begin{eqnarray}\label{actioncab}
{\mathcal L}_{CAB}^{(4)} &=&{T_p \o 2^3 \G(p-2)} I_0 \e^{a _1\cdots a_{p+1}} \Bigg\{ \bigg[  -4 \nabla^a{}_b B_{a _1a _2}\nabla^b{}_a(2\a' F)_{a _3a _4}\cr
&&\cr &&  + 2\nabla^aH_{a _1a _2b}\nabla^b{}_a(2\a'F)_{a _3a _4} -\left( 2\nabla^a{}_bH_{a _1a _2a} -\nabla^a{}_aH_{a _1a _2b}\right)\nabla^b(2\a'F)_{a _3a _4}\cr
&&\cr && -{2\o3}\nabla^{ba}{}_aH_{a _1a _2a _3}(2\a'F)_{ba _4}-{4\o 3}\nabla^a{}_bH_{a _1a _2a _3}\nabla^b(2\a'F)_{aa _4} \bigg] C_{a_5\cdots a_{p+1}}^{(p-3)}\cr
&&\cr && +\left[ 2 \nabla^a{}_iH_{a _1a _2a}(2\a'F)_{a _3a _4} -\nabla^a{}_aH_{a _1a _2i}(2\a'F)_{a _3a _4}\right] \nabla^iC_{a_5\cdots a_{p+1}}^{(p-3)} \cr
&&\cr && +{p-3 \o 3} \nabla^{ia}{}_aH_{a _1a _2a _3}(B+2\a' F)_{a _4a _5}C_{ia_6\cdots a_{p+1}}^{(p-3)} \Bigg\}.\cr &&
\end{eqnarray}
This effective action encodes the corrections to the vertex in fig.\ (3.c). At this moment we have obtained all vertices including $\a'^2$ corrections if present, of the
vertices involved in the diagrams (2.a)-(2.e).  For all field theory vertices except the ones represented in fig.\ (3), the corresponding string amplitudes do not receive higher derivative corrections (the low momentum expressions are exact).  As a result, these other vertices can only possibly be modified by corrections which vanish when the incoming particles are on-shell.  The effect of such modifications (which can be thought of as the leading piece of something proportional to one of the lower order equations of motion) can always be undone by making further modifications to higher-point vertices (so that the full equation of motion appears) as in the discussion in section \ref{subsubsec:CB}.  As such, we can choose to simply leave the vertices uncorrected to begin with.


\subsubsection{$C^{(p-3)}$, $B$, $B$ contact term. }

Using the above results it is possible to compute the field theory result for the diagrams (2.a) -- (2.e). While figs. (2.a) and (2.c) do not receive corrections to
order $\a'^2$, the diagrams in figs. (2.b), (2.d) and (2.e) do receive corrections arising from the corrections to the vertices described in the previous section. It is straightforward
but lengthy to compute the field theory result for (2.a)--(2.e) and we will omit the details here. The field theory result for diagram (2.f) is then obtained by
taking the string theory amplitude, expanding about small momenta to order $\a'^2$ and subtracting the result for diagrams (2.a)--(2.e).
The field theory result
for the diagram in fig.\ (2.f) is lengthy and we will not present it here since it is only used as an intermediate step to obtain the effective D$p$-brane action
which we discuss in detail in the next section.


\section{New four-derivative  D$p$-brane couplings}
\label{sec:HigherDerivativeBraneCouplings}

Given the result for the field theory diagram in fig.\ (2.f) we can now extract the effective action involving
one R-R potential $C^{(p-3)}$ and two $B_{NS}$ fields to order $\a'^2$. Since the results are cumbersome we will perform
all checks possible given the limited set of four-derivative couplings we know. We will require the new couplings to
be invariant under R-R and NS-NS gauge transformations. Moreover, we will require the D$p$-brane action to be compatible with
T-duality.

\subsection{Gauge transformations}

\subsubsection{$B_{NS}$ field gauge transformations}

The field theory result for diagram (2.f) can be obtained from the
following effective action
\begin{equation}
{\cal L}_{CBB} = {\cal L}_{CBB}^{(0)}+ {\cal L}_{CBB}^{(4)},
\end{equation}
where
\begin{equation}
{\mathcal L}^{(0)}_{CBB}= {T_p\o 2^ 3 \G (p-2)} \e^{a_1 \cdots a_{p+1}}  {\tilde B}_{a_1a_2}{\tilde B}_{a_3a_4}C_{a_5\cdots a_{p+1}}^{(p-3)}
\end{equation}
and
\begin{equation}
\begin{split}
 {\mathcal L}_{CBB}^{(4)}  = & {T_p\o 2^ 3 \G (p-2)}{I_0 \o \pi^2}  \e^{a_1 \cdots a_{p+1}}\Bigg[ \bigg( -2\nabla^a{}_b{ B}_{a_1a_2}\nabla^b{}_a{ B}_{a_3a_4}- {1\o 2}\nabla^iH_{a_1a_2a}\nabla_i H_{a_3a_4}{}^a  \cr
& + {1\o 2}\nabla^aH_{a_1a_2i}\nabla_a H_{a_3a_4}{}^i  +H_{a_1a_2i}\nabla^{ia}H_{a_3a_4a} -  {2\o 3}\nabla^iH_{a_1a_2a_3}\nabla^aH_{a_4ai} +{2\o 3}H_{a_1ai}\nabla^{ai}H_{a_2a_3a_4} \cr
& -2\nabla^aH_{a_1a_2a}\nabla^bH_{a_3a_4b}  +2\nabla^aH_{a_1a_2b}\nabla^b{}_a{ B}_{a_3a_4} -2\nabla^a{}_bH_{a_1a_2a}\nabla^b{ B}_{a_3a_4}\cr
& +\nabla^a{}_aH_{a_1a_2b}\nabla^b{ B}_{a_3a_4} -{2\o3}\nabla^{ba}{}_aH_{a_1a_2a_3}{ B}_{ba_4} -{4\o 3}\nabla^a{}_bH_{a_1a_2a_3}\nabla^b{ B}_{aa_4}\bigg)C_{a_5\cdots a_{p+1}}^{(p-3)}\cr
& +\bigg(H_{a_1a_2 i}\nabla^aH_{a_3a_4a}+2\nabla^a{}_iH_{a_1a_2a}{ B}_{a_3a_4}-\nabla^a{}_aH_{a_1a_2i}{ B}_{a_3a_4}\bigg)\nabla^iC_{a_5\cdots a_{p+1}}^{(p-3)}\cr
&-{p-3\o 3} \bigg( \nabla^iH_{a_1a_2a_3}\nabla^bH_{a_4a_5b}- \nabla^{ia}{}_aH_{a_1a_2a_3}{ B}_{a_4a_5}\bigg) C_{ia_6\cdots a_{p+1}}^{(p-3)}\Bigg].
\end{split}
\end{equation}
The sum of ${\cal L}_{CBB}$, ${\cal L}_{CAB}$ in
eqn. (\ref{actioncab}) and ${\cal L}_{CAA}$ in eqn. (\ref{actioncaa})
assembles into ${\cal L}_{C \tilde B \tilde B}$, where
\begin{equation}
\tilde B = B + 2 \a' F,
\end{equation}
{\it i.e.} it has the same form as ${\cal L}_{CBB}$ except $B$ is replaced by
$\tilde B$, which means it is manifestly invariant under gauge transformations
of the $B_{NS}$ field. The overall factor in front of the action was determined
using the coefficient of the zero derivative term. The Lagrangian ${\cal L}_{CB}^{(4)}$ is, of course, invariant
under NS-NS gauge transformations since it depends on $H=dB_{NS}$ only.

\subsubsection{R-R gauge transformations}  \label{sec:Gaugeinvariance}

Next we consider gauge transformations of the R-R potentials, in particular
consider the gauge transformations
\begin{equation}
\begin{split}
& \d C^{(p+1)}=d\L^{(p)}+H\wedge \L^{(p-2)},\\
& \d C^{(p-1)}=d\L^{(p-2)}+ H\wedge \L^{(p-4)},\\
& \d C^{(p-3)}= d\L^{(p-4)},
\end{split}
\end{equation}
which leave the R-R field strength $\tilde F = dC^{(n-1)}+ H \wedge C^{(n-3)}$, invariant.

It turns out that the Lagrangian ${\cal L}_C + {\cal L}_{C \tilde B} + {\cal L}_{C\tilde B \tilde B}$ changes after performing
R-R gauge transformations by a quantity which vanishes on-shell. Here
\begin{equation}
{\mathcal L_C}={T_p\o \G(p+2) }\e ^{a_1\cdots
a_{p+1}}C^{(p+1)}_{a_1\cdots a_{p+1}}.
\end{equation}
It is possible to use the ambiguity of adding terms which vanish on-shell to obtain an effective Lagrangian
which is invariant under R-R gauge transformations. The terms which need to be added are
\begin{equation}
\Delta{\cal L}_{CB}=-{T_p\o 2^2 \G(p) }{I_0\o
\pi^2}\e ^{a _1\cdots a_{p+1}}\nabla^{a}H_{a _1a _2a}\nabla^\m{}_\m
C^{(p-1)}_{a_3\cdots a_{p+1}}
\end{equation}
which leads to the following correction of the $CBB$ coupling
\begin{equation}
\begin{split}
\D{\mathcal L}_{CBB}=& {T_p\o 2^ 3 \G(p-2)}{I_0\o \pi^2}\e ^{a _1 \cdots a_{p+1}} \bigg[  -\nabla^aH_{a _1a _2 a}H_{a _3a _4 i}\nabla^i C_{ a_5 \cdots a_{p+1}}\cr
& +\left( \nabla^a{}_bH_{a _1a _2 a}H_{a _3a _4 b}  +2 \nabla^aH_{a _1a _2 a}\nabla^bH_{a _3a _4 b}  -{2\o 3} \nabla^aH_{a _1b a}\nabla^b H_{a _2a _3a _4} \right) C_{ a_5 \cdots a_{p+1}}         \cr
& + {p-3 \o 3} \nabla^aH_{a _1a _2 a}\nabla^i H_{a _3a _4a _5} C_{i\,a_6 \cdots a_{p+1}}\bigg].
\end{split}
\end{equation}

\subsubsection{The new four-derivative couplings}

The Lagrangian which is invariant under NS-NS and R-R gauge transformations is
\begin{equation} \label{final}
{\cal L} = {\cal L}_C + {\cal L}_{C\tilde B} + {\mathcal L}_{C\tilde B \tilde B } +\D  {\cal L}_{C B} + \D{\mathcal L}_{CB B },
\end{equation}
where
\begin{equation}\label{eq:ActionBC}
\begin{split}
 {\cal L}_{CB}+\D {\cal L}_{CB} = &  {T_p \o 2 \G(p)}\e ^{ a _1 \cdots a_{p+1} }  \bigg[ B_{ a _1 a _2}C^{(p-1)}_{ a_3 \cdots a_{p+1} }  + {I_0\o 2 \pi^2} \big(  -\nabla^a{}_aH_{a_1 a _2i}\nabla^iC^{(p-1)}_{ a_3 \cdots a_{p+1} }\cr
 & + 2 \nabla^a{}_iH_{ a _1 a _2a}\nabla^iC^{(p-1)}_{ a_3 \cdots a_{p+1} }+{p-1 \o 3 } \nabla^{ia}{}_aH_{ a _1 a _2 a _3}C^{(p-1)}_{i\, a_4 \cdots a_{p+1} }\big)\bigg] ,
\end{split}
\end{equation}
and
\begin{equation}
\label{eq:FinalCBB}
\begin{split}
& {\mathcal L}_{C\tilde B\tilde B}+\D{\cal L}_{CBB} \cr
& = {T_p\o 2^ 3 \G (p-2)} \e^{a_1 \cdots a_{p+1}}  {\tilde B}_{a_1a_2}{\tilde B}_{a_3a_4}C_{a_5\cdots a_{p+1}}^{(p-3)}+ {T_p\o 2^3 \G(p-2) }{I_0\o \pi^2}\e^{a_1\cdots a_{p+1}}\Bigg[ \bigg( -2\nabla^a{}_b{\tilde B}_{a_1a_2}\nabla^b{}_a{\tilde B}_{a_3a_4} \cr
&  -{1\o 2}\nabla^iH_{a_1a_2a}\nabla_i H_{a_3a_4}{}^a  + {1\o 2}\nabla^aH_{a_1a_2i}\nabla_a H_{a_3a_4}{}^i+ H_{a_1a_2\m}\nabla^{\m a}H_{a_3a_4a}- {2\o 3}\nabla^\m H_{a_1a_2a_3}\nabla^aH_{a_4a\m}   \cr
& + {2\o 3}H_{a_1ai}\nabla^{ai}H_{a_2a_3a_4} + 2\nabla^aH_{a_1a_2b}\nabla^b{}_a{\tilde B}_{a_3a_4}-  2\nabla^a{}_bH_{a_1a_2a}\nabla^b{\tilde B}_{a_3a_4}  + \nabla^a{}_aH_{a_1a_2b}\nabla^b{\tilde B}_{a_3a_4} \cr
& - {2\o3}\nabla^{ba}{}_aH_{a_1a_2a_3}{\tilde B}_{ba_4} - {4\o 3}\nabla^a{}_bH_{a_1a_2a_3}\nabla^b{\tilde B}_{aa_4}\bigg)C_{a_5\cdots a_{p+1}}^{(p-3)} +{p-3 \over 3 } \nabla^{ia}{}_aH_{a_1a_2a_3}{\tilde B}_{a_4a_5}C_{ia_6\cdots a_{p+1}}^{(p-3)} \cr
& +\bigg( 2 \nabla^a{}_iH_{a_1a_2a}{\tilde B}_{a_3a_4} - \nabla^a{}_aH_{a_1a_2i}{\tilde B}_{a_3a_4}\bigg)  \nabla^iC_{a_5\cdots a_{p+1}}^{(p-3)} \Bigg].
\end{split}
\end{equation}

Because of the gauge invariance, the four-derivative part of this action can be rewritten in terms of the R-R field strength rather than the potential, giving
\begin{multline}
\mathcal{L}^{(4)}=\frac{T_pI_0}{4\pi^2}\e^{a_1\cdots a_{p+1}}\left\{\lp-\nabla^bH_{a_1b}^{\hphantom{a_1b}i}H_{a_2a_3i}-\hlf\nabla^bH_{a_1a_2}^{\hphantom{a_1a_2}i}H_{a_3bi}-\hlf\nabla^iH_{a_1a_2}^{\hphantom{a_1a_2}b}H_{a_3bi}\right.\right.\\
\left.\left.+\nabla^{b\hphantom{b}c}_{\hphantom{b}b}\widetilde{B}_{a_1a_2}\widetilde{B}_{a_3c}+2\nabla^{bc}\widetilde{B}_{a_1a_2}\nabla_b\widetilde{B}_{a_3c}+\nabla^b_{\hphantom{b}b}\widetilde{B}_{a_1}^{\hphantom{a_1}c}\nabla_c\widetilde{B}_{a_2a_3}\right.\right.\\
\left.\left.+2\nabla_{a_1}^{\hphantom{a_1}bc}\widetilde{B}_{a_2b}\widetilde{B}_{a_3c}-\nabla_{a_1}^{\hphantom{a_1}b}B_{a_2a_3}\nabla^c\widetilde{B}_{bc}-\nabla^{bc}\widetilde{B}_{a_1b}\nabla_c\widetilde{B}_{a_2a_3}\rp\frac{1}{\G(p-1)}F^{(p-2)}_{a_4\cdots a_{p+1}}\right.\\
\left.+\lp\nabla^{bi}H_{a_1a_2b}-\hlf\nabla^b_{\hphantom{b}b}H_{a_1a_2}^{\hphantom{a_1a_2}i}\rp\ls\frac{2}{\G(p)}\widetilde{F}^{(p)}_{ia_3\cdots a_{p+1}}+\frac{1}{\G(p-2)}\widetilde{B}_{a_3a_4}F^{(p-2)}_{ia_5\cdots a_{p+1}}\rs\right\}.
\end{multline}
Here $F^{(p-2)}=dC^{(p-3)}$.  Note that the terms in the third line above all involve the combination $\nabla^b\widetilde{B}_{bc}$, which is the leading term in the equation of motion for the gauge field and hence can be removed by a field redefinition\footnote{The equation of motion for the gauge fields is given by
\be
0=\nabla^b\widetilde{B}_{bc}+\cdots,
\ee
where $\cdots$ includes both higher derivative terms as well as terms with the same number of derivatives but at least one R-R potential.  Thus we can remove the terms in question by a field redefinition, at the cost of introducing new terms with two or more R-R fields.}.  After removing such terms, the action (\ref{eq:FinalResult}) that remains is the main result of this paper.

In the section \ref{subsec:TDuality}, we will perform a consistency check of this new action. We will check that it is compatible with T-duality
at the linearized level.

\subsection{Compatibility with T-duality}
\label{subsec:TDuality}

Given the Lagrangian on a D$p$-brane world-volume, T-duality will in general mix terms with different numbers of fields and the same
number of derivatives. Schematically we have represented how T-duality acts on the four-derivative terms in the Lagrangian in fig.\ 7. Under T-duality
we expect the complete Lagrangian to map to itself. But we do not know the entire set of four-derivative couplings yet.
In this paper we have determined
the four-derivative couplings involving $C^{(p-1)},B$ and $C^{(p-3)} B, B$ while the couplings involving $C^{(p-3)} h, h$ were already known.
As illustrated in fig.\ 7 given the couplings we know one consistency check is to apply T-duality along the brane and the $C^{(p-3)} B, B$ couplings
then map to themselves. In fig.\ 7 we have used the fact that there are no terms in the brane Lagrangian involving one NS-NS field and one R-R potential of degree
$n \geq p+5$, $p\leq p-3$, or terms involving 2 NS-NS fields and one R-R potential of degree $n \leq p-5$ or $n\geq p+7$.

Lets consider a space-time with a $U(1)$ isometry in a direction labeled by $y$ and a brane positioned so that
$y$ is parallel to it. The Lagrangian  ${\cal L}$ (in eqn. \ref{action1}) together with  ${\cal L}_{Chh}$ which involves the metric $h$ is the Hodge dual
with respect to the brane coordinates of
\begin{equation}
X = C^{(p-3)} \wedge X^{(4)} +C_i^{(p-3)} \wedge X^{(5) i} + \nabla_i C^{(p-3)} \wedge Y^{(5)i},
\end{equation}
where
\begin{equation}
C_i^{(p-3)} = C_{i a_1 \dots a_{p-2}} dx^{a_1 \cdots a_{p-2}},
\end{equation}
and the forms $X^{(4)}$, $X^{(5)i}$ and $Y^{(5)i}$ can be obtained from the Lagrangian.

A D$p$-brane action is expected to be compatible with T-duality. To check this we first drop
all derivatives with respect to the coordinate $y$. Lets label the form which is obtained from $X$
after applying T-duality by $X'$. Compatibility with T-duality then translates into
\begin{equation}
X=X',
\end{equation}
which, as explained above, can only be verified at the linearized level. It is not difficult to see that the new couplings do indeed map to
themselves under T-duality to leading order in the number of fields.

Lets consider T-duality in a direction transverse to the brane. Even though there are no covariant four-derivative interactions involving $C^{(p-5)}$, $B$, $h$, there can be such an interaction if one direction is singled out, if for example one direction is an isometry.  Such non-covariant terms could be generated by applying T-duality transverse to the brane to the interaction involving $C^{(p-3)}$ and two NS-NS fields.  It is easy to see from the form of the Lagrangian that such terms will not be generated.

Once the entire four-derivative action is known it should be possible
to check compatibility with T-duality to all orders in the fields \cite{progress}.
\begin{figure}[!htb]
\label{sugracab}
\begin{center}
\scalebox{.5}{\includegraphics{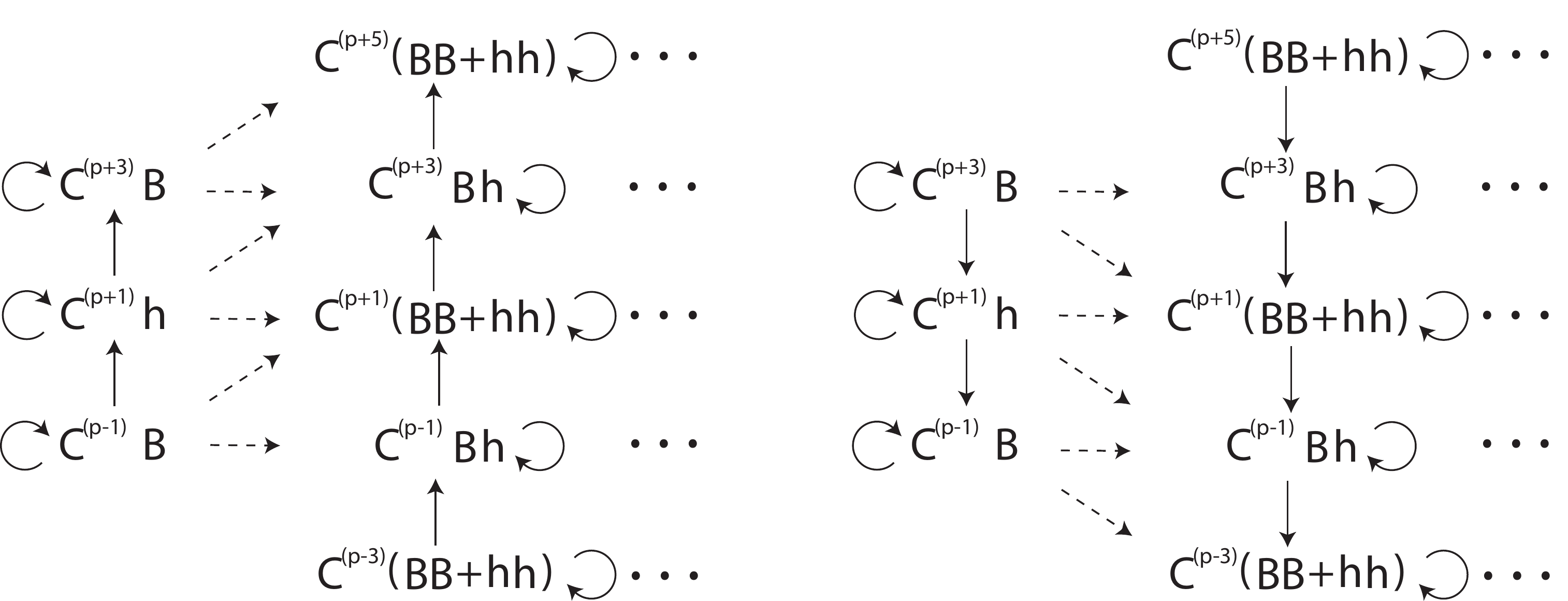}} 
\caption{This diagram sketches the action of T-duality. Each entry represents the fields involved in a certain four-derivative
term in the Lagrangian.
 The solid arrows show the action of the T-duality rules to leading order in the number of fields.
The dashed lines show the first non-linear contributions.
The first diagram represents the action of
a T-duality along the brane and the second transverse to the brane. }
\end{center}
\end{figure}

\subsection{Comparison with existing literature}

The last term in the first line of (\ref{eq:FinalCBB}) is the only term that is quadratic in gauge fields, and was previously determined in ref. \cite{Wyllard:2000qe} and \cite{Wyllard:2001ye}; we
have agreement with their results. The next two terms represent the gauge invariant completion of the interactions involving $C^{(p-3)}$ presented in ref. \cite{Becker:2010ij}, and so this action is consistent with that result as well.

The terms in (\ref{eq:ActionBC}) are consistent with \cite{Garousi:2010ki} up to terms which vanish on-shell (specifically, a term proportional to $\nabla^\m H_{ai\m}$).  In \cite{Garousi:2010bm,Garousi:2011ut}, the authors follow a similar procedure to the one we have here, but to facilitate the comparison to field theory they restrict their analysis to the situation in which $(p_2p_3)=(p_2Dp_3)=0$.  As such they miss some terms in which derivatives on each $B$ field contract with each other, including the term quadratic in gauge fields mentioned above.  However, for the subset of terms which they find, the coefficients agree, up to some signs.

The results presented in (\ref{eq:FinalResult}) are the first time that the complete set of gauge completed four-derivative corrections involving $C^{(p-3)}$ and two $B$ fields has appeared in the literature, and it is in agreement with all partial results that have appeared previously.

\section{Conclusion}

String theory is a theory of quantum gravity as opposed to merely a theory of classical gravity. As such it is
interesting to compute corrections, either $\a'$ or $g_s$ corrections. In this paper we have  computed contributions
to the D$p$-brane action of order $\a'^2$, which compared to the $\a'^3$ corrections arising from the low energy effective
action of type II theories in ten dimensions are dominant. Higher derivative terms in the Lagrangian can usually be neglected. The reason
they can be relevant in flux backgrounds, for example, is that they modify equations of motion and higher derivative
contributions can become of order 1 if it is integrated over a higher dimensional space.  This is known to happen for the gravitational $C^{(4)}\tr(R^2)$ couplings on D7-branes in the context of type IIB flux compactifications.  In \cite{progress} we will compute the four-derivative couplings between $C^{(p-1)}$, $B_{NS}$, and a graviton, and such couplings on D6-branes are expected to play a similar role in resolving a puzzle about the consistency of a class of IIA flux compactifications related to M-theory compactifications on $G_2$ manifolds with flux. It would further be interesting to see if the relation between string amplitudes involving closed strings and open strings found in ref. \cite{Stieberger:2009hq} can be used 
to facilitate the construction of the entire four-derivative D$p$-brane action. 

It is expected that these corrections will modify the supersymmetry conditions and equations
of motion of string theory solutions. It would be interesting to work out concrete examples to see if the
$\a'^2$ corrections lead to small modifications or if they can actually represent new obstructions to
the solvability of the equations of motion.

We have considered amplitudes in which the topology of the srting world-sheet is the disc and insertions of
open or closed string vertex operators. In general, the Euler characteristic of the world-sheet is determined by the number
of boundaries, $b$, the number of cross-caps $c$ and the number of handles $h$ according to the
\begin{equation}
\chi = 2 - b-c-2h.
\end{equation}
We have considered the disc with $\chi=1$.
It would be interesting to further extend the tools we used to obtain subleading terms in the $g_s$ expansion.
The first subleading contribution arises from the annulus diagram with $\chi=0$, while diagrams with at least one boundary
and one handle would have an Euler character $\chi\leq -1$ and are further suppressed. Work in this direction is in progress.

\acknowledgments

The authors would like to thank N.~Berkovits, M.~Green, I.~Klebanov, R.~Minasian, and M.~Rocek for useful discussions and comments. K.B. would like to
thank the Aspen Center for Physics and the IAS Princeton for hospitality and support during the
completion of this work. D.R. would like to thank the IAS for hospitality This
research was supported in part by NSF Grant No. PHY05-55575, NSF Grant No.
PHY09-06222, NSF Gant No. PHY05-51164, Focused Research Grant DMS-0854930,
Texas A\&M University, and the Mitchell Institute for Fundamental Physics and Astronomy.


\appendix
\section{Some Integrals}
\label{AppendixA}

In this appendix we will present details about the evaluation of the complex integrals used in the main part of the paper. We
start defining the integrals
\begin{equation}
I_{a,b,c,d,e,f} = \int_{|z_i| \leq 1}  d^2 z_2 d^2 z_3 \tilde {\cal K} {\cal K},
\end{equation}
where
\begin{equation}
\tilde {\cal K} = |z_2|^{2 a}  | z_3 | ^{2 b} (1-| z_2|^2 )^ c (1-|z_3 |^2 )^d | z_2 - z_3 | ^{ 2 e} |1- z_2 \bar z_3 |^{ 2 f},
\end{equation}
and
\begin{equation}
{\cal K}=|z_2| ^{2p_1\cdot p_2}|z_3|
^{2p_1\cdot p_3} \left(1-|z_2|^2 \right)^{p_2\cdot Dp_2} \left(1-|z_3|^2\right)^{p_3\cdot Dp_3}| z_2-z_3| ^{2p_2\cdot p_3}| 1-z_2\bar
z_3| ^{2p_2\cdot Dp_3}.
\end{equation}
Note that ${\tilde K}$, has the same form as ${\cal K}$ but different exponents. In general, $a, b, \dots$ are some positive or negative integers.
We expect $I_{a,b,c,d,e,f}$ to be a meromorphic functions of $p_i \cdot p_j$ and $p_k \cdot Dp_l$, for any $i,j,k,l$ which we view as several complex variables.
The above integral defines
this function for large enough momenta while in other regions it has to be defined using analytic continuation. We are interested in Laurent expansion
of $I$ close to zero, and in particular in terms of $O(p^{-4})$ and $O(p^0)$.

We start introducing polar coordinates $z_i = r_i e^{ i \th_i}$, $i=2,3$. Since the integrant depends only on $\th_2-\th_3$ one of the integrals can be
explicitly performed. Next we Taylor expand the integral in $r_2$ and $r_3$ using
\begin{equation}
{1\o (1-x)^s}=\sum_{n=0}^\infty \left(
 \begin{array}{c}
 s+n-1 \\
n \\
 \end{array}
 \right)x^n
\end{equation}
for $\mid x\mid \leq 1$, carefully separating the regions in which $r_2\leq r_3$ and $r_3 \leq r_2$. The integrals over the radial coordinates
are then easy to perform and the result is a set of infinite sums
\begin{equation}\label{appendix1}
\begin{split}
 \pi^2 \sum_{n_i = 0 }^ \infty  & \left(   {1\over p_1 \cdot p_2 + 1 + a + n_1+ n_3+ n_5 }+{1\over p_1 \cdot p_3 + 1 + b + n_2+ n_3+ n_5 } \right)  \cr
& {1\over p^2 + 2+ a+ b + e+n_1 +n_2 +n_5 + n_6 } \d_{n_3-n_4+n_5-n_6,0}  \cr
& \begin{pmatrix} -p_2 Dp_2 -1 - c +n_1 \\ n_1 \end{pmatrix}  \begin{pmatrix} -p_3 Dp_3 -1 - d +n_2 \\ n_2 \end{pmatrix} \\
&  \begin{pmatrix} -p_2 p_3 -1 - e +n_3 \\ n_3 \end{pmatrix}  \begin{pmatrix} -p_2 p_3 -1 - e +n_4 \\ n_4 \end{pmatrix}\\
&  \begin{pmatrix} -p_2 D p_3 -1 - f +n_5 \\ n_5 \end{pmatrix}  \begin{pmatrix} -p_2 D p_3 -1 - f +n_6 \\ n_6 \end{pmatrix},
\end{split}
\end{equation}
where the Kronecker delta symbol arises from the integration over $\th_2-\th_3$.

Next we define the following integrals which are enough to evaluate the three-point function
\begin{eqnarray}
I_0&=& \int_{|z_i|\leq 1}
d^2z_2d^2z_3{\left(z_3\bar z_2-z_2\bar z_3\right)^2\o 2\mid z_2\mid
^2\mid z_3\mid ^2\mid 1-z_2\bar z_3\mid ^2\mid z_2-z_3\mid ^2}{\cal
K}\cr &&\cr I_1&=& \int_{|z_i|\leq 1} d^2z_2d^2z_3{|1+z_2\bar
z_3|^2\o |1-z_2\bar z_3|^2|z_2|^2|z_3|^2}{\cal K}\cr &&\cr
I_2&=&\int_{|z_i|\leq 1} d^2z_2d^2z_3{|z_2+z_3|^2\o |z_2-
z_3|^2|z_2|^2|z_3|^2}{\cal K}\cr &&\cr I_3 &=& \int_{|z_i|\leq 1}
d^2z_2d^2z_3{(1+|z_2|^2)(1+|z_3|^2)\o
(1-|z_2|^2)(1-|z_3|^2)|z_2|^2|z_3|^2}{\cal K}\cr && \cr I_4&=&
\int_{|z_i|\leq 1} d^2z_2d^2z_3{2(1+|z_2|^2)\o
|z_2|^2|z_3|^2(1-|z_2|^2)}{\cal K}\cr &&\cr I_5&=&
\int_{|z_i|\leq 1} d^2z_2d^2z_3{1-|z_2|^2|z_3|^2\o
|z_2|^2|z_3|^2|1-z_2\bar z_3|^2}{\cal K}\cr &&\cr I_6&=&
\int_{|z_i|\leq 1} d^2z_2d^2z_3{-(1+|z_2|^2)(|z_2|^2-|z_3|^2)\o
|z_2|^2|z_3|^2(1-|z_2|^2)|z_2-z_3|^2}{\cal K}\cr &&\cr I_7&=&
\int_{|z_i|\leq 1} d^2z_2d^2z_3{-(1+|z_2|^2)(1-|z_2|^2|z_3|^2)\o
|z_2|^2|z_3|^2|1-z_2\bar z_3|^2(1-|z_2|^2)}{\cal K}\cr &&\cr
I_8&=&\int_{|z_i|\leq 1}
d^2z_2d^2z_3{(|z_2|^2-|z_3|^2)(1-|z_2|^2|z_3|^2)\o
|z_2|^2|z_3|^2|1-z_2\bar z_3|^2 |z_2-z_3|^2}{\cal K}\cr &&\cr
I_9&=&\int_{|z_i|\leq 1} d^2z_2d^2z_3{(|z_2|^2-|z_3|^2)\o
|z_2|^2|z_3|^2 |z_2-z_3|^2}{\cal K}\cr &&\cr I_{10}&=&
\int_{|z_i|\leq 1} d^2z_2d^2z_3{1\o |z_2|^2|z_3|^2}{\cal K}.
\end{eqnarray}
Using \C{appendix1} it is possible to show that asymptotically in the region of small
momenta the following expansions hold up to terms quadratic in momenta
\begin{eqnarray}
I_0&=& -{\pi^4\o 3}\cr &&\cr I_1&=&I_{10}+{2\o 3}\pi^4\cr &&\cr I_2
&=&I_{10}+{2\pi^2\o p^2(p_2\cdot p_3)}Q_3+{2\pi^2\o p^2(p_2\cdot
p_3)}Q_2 -{2\o 3}\pi^4 -{1\o 3}\pi^4[p_2Dp_2+p_3Dp_3+4p_2Dp_3]{1\o
p_2\cdot p_3}\cr &&\cr I_3 &=&I_{10} +{2\pi^2\o (p_1\cdot
p_2)(p_3Dp_3)}Q_3+{4\pi^2\o (p_2Dp_2)(p_3Dp_3)}[1+{\pi^2\o
6}(p_2Dp_3+p_2\cdot p_3)^2]\cr &&\cr &&  +{2\pi^2\o (p_1\cdot p_3)
(p_2Dp_2)}Q_2-{1\o 3}\pi^4[{p_2D p_2\o p_3D p_3}+{p_3Dp_3\o
p_2Dp_2}+2{p_1\cdot p_3\o p_2D p_2}+2{p_1\cdot p_2\o p_3D p_3}]+{2\o
3}\pi^4\cr && \cr I_4&=& 2I_{10}+{4\pi^2\o (p_1\cdot
p_3)(p_2Dp_2)}Q_2-{2\o 3}\pi^4{p_3Dp_3\o p_2Dp_2}+{2\o 3}\pi^4\cr
&&\cr I_5&=&I_{10}\cr &&\cr I_6&=&-I_9-{2\pi^2\o (p_1\cdot
p_3)(p_2Dp_2)}Q_2+{\pi^4\o3}{p_3Dp_3\o p_2Dp_2}+{2\pi^4\o3}{p_2\cdot
p_3\o p_2Dp_2}+{2\pi^4\o3}{p_2Dp_3\o p_2Dp_2}+{1\o 3}\pi^4 \cr &&\cr
I_7&=&-I_{10}-{2\pi^2\o (p_1\cdot p_3)(p_2Dp_2)}Q_2+{\pi^4\o
3}{p_3Dp_3\o p_2Dp_2}+{2\pi^4\o 3}{p_2\cdot p_3\o p_2Dp_2}+{2\pi^4\o
3}{p_2Dp_3\o p_2Dp_2}-{1\o 3}\pi^4\cr &&\cr I_8&=& I_9\cr &&\cr
I_9&=& {\pi^2\o p^2(p_1\cdot p_3)}Q_2-{\pi^2\o p^2(p_1\cdot
p_2)}Q_3\cr && \cr I_{10}&=&{\pi^2\o p^2(p_1\cdot p_3)}Q_2+{\pi^2\o
p^2(p_1\cdot p_2)}Q_3\cr &&
\end{eqnarray}
where we have used the notation
\begin{eqnarray}
p^2&=&p_1\cdot p_2+p_1\cdot p_3+p_2\cdot p_3\cr Q_2&=& [1-{\pi^2\o
6}p^2(p_2Dp_2)]\cr Q_3&=& [1-{\pi^2\o 6}p^2(p_3Dp_3)],
\end{eqnarray}

For most integrals, like for example $I_{10}$, the momentum expansion can be done before doing the sums.
In this case it is easy to obtain the result since most contributions for $n_i \neq 0$ are of higher
orders in the momentum expansion and the sum localizes at $n_i=0$. However, some situations require more care like the case $e=-1$. In this case for small $p_2 \cdot p_3$ the largest contribution to the sum arises
from large $n_3, n_4$ and these sums have to be done exactly. To evaluate these sums the following
results which hold for $x,y \approx 0$ are useful
\begin{equation}
\begin{split}
& \sum_{n=1}^\infty \begin{pmatrix} -x+n \\ n \end{pmatrix} {1\over y+n} = -{1\over y} +{\G(x) \G(y)\over \G(x+y)}  , \cr
& \sum_{n=1}^\infty \begin{pmatrix} -x+n \\ n \end{pmatrix}^2 {1\over y+n} = {1\over 2 x}-{\pi^2 \over 4} x -{\pi^2 \o 6}y +\dots , \cr
& \sum_{n=1}^\infty \begin{pmatrix} -x+n \\ n \end{pmatrix}^2 {1\over y+1+n} = {1\over 2 x}-1 +{\pi^2 \over 12} x +\left( 1- {\pi^2 \o 6}\right) y +\dots , \cr
& \sum_{n=1}^\infty \begin{pmatrix} - x + n \\ n \end{pmatrix} \begin{pmatrix} - x +1+  n \\ 1+n \end{pmatrix}{1\over y+1+n} =
{1\o 2 x} - 1 + \left(1-{\pi^2 \o 12}  \right) x + \left(1-{\pi^2 \o 6} \right) y + \dots
\end{split}
\end{equation}
The first sum can be done exactly while the result for the next sums is quoted only up to terms quadratic in $x$ and $y$.

\end{document}